\documentclass[sigconf]{acmart}
\AtBeginDocument{%
  }

\copyrightyear{2026}
\acmYear{2026}
\setcopyright{cc}
\setcctype{by}
\acmConference[CHI '26]{Proceedings of the 2026 CHI Conference on Human Factors in Computing Systems}{April 13--17, 2026}{Barcelona, Spain}
\acmBooktitle{Proceedings of the 2026 CHI Conference on Human Factors in Computing Systems (CHI '26), April 13--17, 2026, Barcelona, Spain}
\acmPrice{}
\acmDOI{10.1145/3772318.3790571}
\acmISBN{979-8-4007-2278-3/2026/04}

\usepackage{makecell}

\usepackage{color}
\newcommand{\general}[1]{{\textcolor{black}{#1}}}

\newcommand{\name}{ASafePlace}
\begin{document}

\title{ASafePlace: User-Led Personalization of VR Relaxation via an Art Therapy Activity}

\author{Chuyang Zhang}
\authornote{Both authors contributed equally to this research.}
\orcid{0009-0009-7088-9953}
\affiliation{%
  \department{School of Design}
  \institution{Southern University of Science and Technology}
  \city{Shenzhen}
  \country{China}
}
\affiliation{%
  \institution{University of Calgary}
  \city{Calgary}
  \country{Canada}
}
\email{chuyang.zhang1@ucalgary.ca}

\author{Bin Yu}
\authornotemark[1]
\orcid{0000-0002-3128-7441}
\affiliation{%
  \institution{Amsterdam University of Applied Science}
  \city{Amsterdam}
  \country{Netherlands}}
\email{b.yu@hva.nl}

\author{Yuchao Wang}
\orcid{0009-0004-9017-6663}
\affiliation{%
  \department{School of Design}
  \institution{Southern University of Science and Technology}
  \city{Shenzhen}
  \country{China}
}
\email{12310531@mail.sustech.edu.cn}

\author{Mansi Yuan}
\orcid{0009-0002-6289-3820}
\affiliation{%
 \institution{Dominican University of California}
 \city{San Rafael}
 \state{California}
 \country{United States}}
\email{mansi.yuan@students.dominican.edu}

\author{Wanqi Wang}
\orcid{0009-0005-9469-7797}
\affiliation{%
  \institution{Wuhan University}
  \city{Wuhan}
  \country{China}}
\email{wwq202303@163.com}

\author{Seungwoo Je}
\orcid{000-0002-3968-7298}
\affiliation{%
  \institution{Southern University of Science and Technology}
  \city{Shenzhen}
  \country{China}}
\email{seungwoo@sustech.edu.cn}

\author{Pengcheng An}
\authornote{Corresponding author.}
\orcid{0000-0002-7705-2031}
\affiliation{%
  \department{School of Design}
  \institution{Southern University of Science and Technology}
  \city{Shenzhen}
  \country{China}}
\email{anpc@sustech.edu.cn}




\begin{abstract}

\general{To overcome the lack of deep personalization in standard biofeedback methods, we introduce ASafePlace, a system utilizing an AI-powered, \general{art-therapy-inspired} exercise called \textit{The Safe Place}, to create a personalized VR biofeedback experience.} In our system, users sketch a personal sanctuary from memory, which is then transformed into a customized 360° virtual environment with personalized audio guidance for relaxation training. A study with 52 participants showed this approach effectively reduced anxiety and increased user presence, while the integration of \general{art-therapy-inspired activity} and biofeedback produced strong improvements in physiological relaxation, measured by heart rate variability and respiration rate. Qualitative results showed how participants' sense of familiarity and presence was enhanced by the symbolic elements and natural sanctuaries created from their autobiographical memories. Our findings demonstrate that \general{art-therapy-inspired activity} is a powerful tool for creating highly effective and individualized relaxation experiences, naturally connecting the virtual environment to a user's core memories and emotions.

\end{abstract}


\begin{teaserfigure}
  \includegraphics[width=\textwidth]{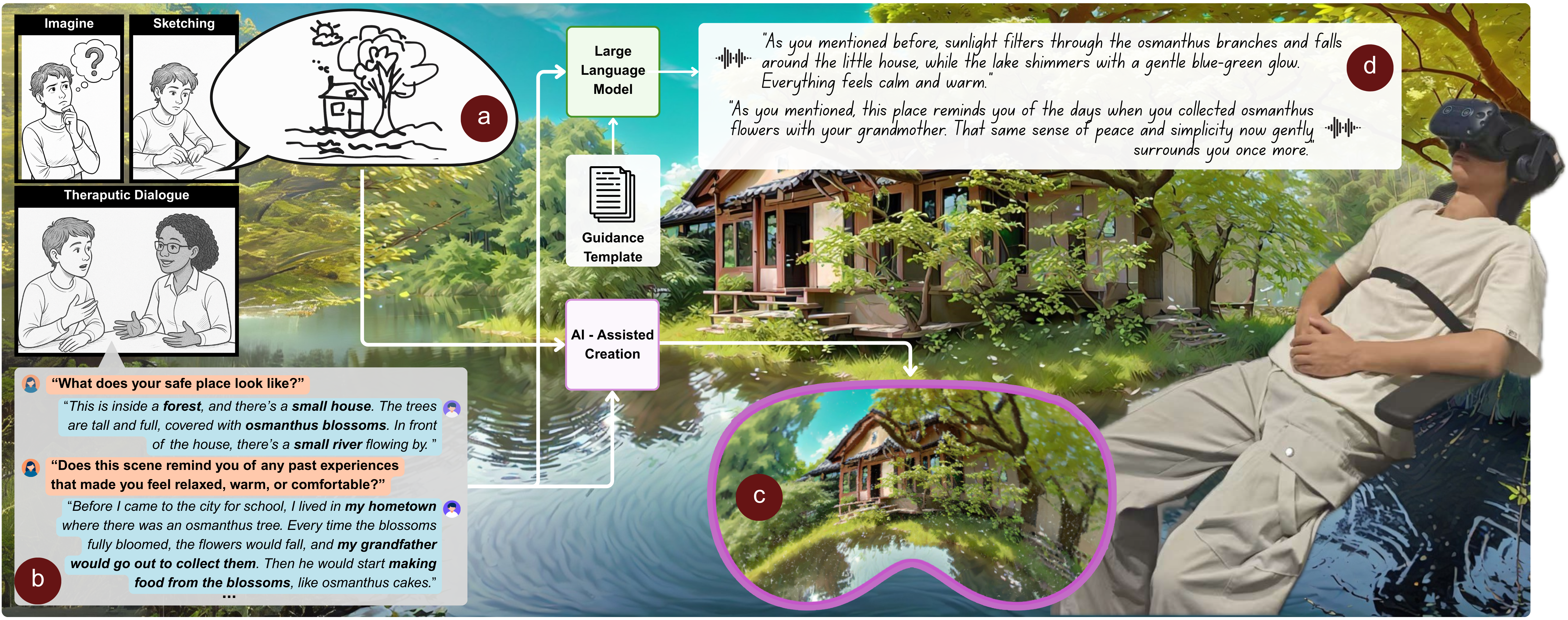}
  \caption{The ASafePlace system enables users to relax in an AI-assisted, user-authored 360° virtual environment with audio guidance infused with their personal life experiences, derived from an \general{art-therapy-inspired activity}. }
  \Description[A diagram of the system pipeline from co-creation to a personalized 360° virtual sanctuary]{On the left, the top row shows small illustrations of a user imagining, sketching, and verbally describing a safe place with an art therapist. Next to these icons, Figure 1(a) shows an example user sketch: a simple scene with a house and a tree beside the water. Below, Figure 1(b) shows an example text conversation in which the therapist asks what the safe place looks like and whether it connects to warm or comforting memories. The user explains that the safe place is a small house in a forest with osmanthus trees, and recalls childhood experiences of his grandfather collecting osmanthus blossoms and making cakes. Arrows indicate that the sketch and conversation are used as inputs to generate two outputs: Figure 1(c), a 360° environment map depicting a house in a forest, and Figure 1(d), audio guidance that references the user’s childhood memory and associated feelings. The background of the figure is the full 360° environment map. On the right, an illustration shows the user relaxing while immersed in the 360° virtual environment.}
  \label{fig:teaser}
\end{teaserfigure}


\maketitle

\section{Introduction}





Biofeedback-based relaxation training is a well-established method for alleviating stress and enhancing mental well-being \cite{2, 8, 46, 47}. In biofeedback systems, users' physiological data are measured and typically presented through graph-based visualizations or abstract interfaces such as dashboards \cite{whited2014effectiveness} and mobile applications \cite{minen2021heartrate, son2023use}. Though effective in supporting self-regulation \cite{kennedy2019biofeedback}, these biofeedback systems may also increase cognitive load \cite{yu2018delight}, trigger 'relaxation-induced anxiety' \cite{yu2018biofeedback} or result in low engagement in long term use \cite{wollmann2016user}. In relaxation training, biofeedback interfaces serve both as information carriers and as sensory stimuli shaping the user experience. Balancing the effectiveness of biofeedback for self-regulation with the overall experience of relaxation training has become an important challenge increasingly explored in HCI research. In particular, Virtual Reality (VR)-based biofeedback approaches have been widely shown to enhance both relaxation effectiveness and the meditative experience. VR offers unique advantages for relaxation training \cite{12,13,14,15} by creating immersive, multi-sensory environments that help users disengage from external distractions and focus on relaxation \cite{1,3,16,17,18}. 


Individuals vary widely in how they respond to relaxation interventions (e.g., baseline traits and prior experience \cite{430}, personality and dispositional traits \cite{431}), and such individual differences may lead to distinct preferences for relaxation environments. As Pardini et al. \cite{25} demonstrate, offering users choices, such as sea, mountain, or countryside contexts as well as atmospheric variables, leads to stronger enjoyment and enhanced relaxation. Similarly, Gao et al. \cite{432} found substantial variation in environments preferences for mood improvements. Moreover, Cheng et al. \cite{433} observed that individuals with higher vividness of mental imagery reported stronger presence in virtual spaces, suggesting that cognitive traits also affect the experience of immersion. Personalization is therefore considered an important route to increase engagement and therapeutic benefit in relaxation tools, including VR biofeedback systems. However, personalization of current systems still mostly rely on superficial configuration options or pre-built fixed scenarios. A review study \cite{435} further underscores the inconsistent or limited personalization in deployed apps and call for more rigorous, scalable personalization methods.

Personalization was highlighted in prior studies \cite{25, 31, 32} as crucial for improving the effectiveness of relaxation training. For instance, allowing individuals to adapt the visual or auditory features of a relaxation exercise can make the experience feel more meaningful and supportive of their emotional state \cite{33}.  Personalizing user interfaces to individual needs and preferences can create context-aware experiences, improving user satisfaction and engagement \cite{de2025sonora}. Current personalization approaches typically depend on structured user input or rule-based configuration to adjust interfaces and system settings, such as VR environments \cite{19, 31, 32}, audio tracks \cite{31, 32, 34}, haptic feedback \cite{34}, virtual avatars \cite{29} or game difficulty and narratives \cite{al2014dodging}. Although these studies allow for a degree of customization, many highlight the need for deeper personalization to support more viable and sustainable relaxation experiences. For instance, incorporating a person’s lived experiences and memories into relaxation training may create strong emotional associations \cite{31} and personal resonance \cite{stelter2009experiencing}, which can enhance engagement and the overall effectiveness of the intervention.

Art therapy (AT) \cite{reynolds2000effectiveness} is a highly personalized therapeutic practice that has been widely applied in relaxation and anxiety reduction, often in combination with other techniques such as paced breathing \cite{kim2014effects}, progressive muscle relaxation \cite{liu2022expressive}, and mindfulness \cite{jang2016beneficial}. It supports self-expression and disclosure of emotions, memories, and lived experiences through art-making \cite{tang2019art}. AT also involves verbal reflection, for instance, talking about their artwork, describing feelings and thoughts evoked by the process. Although art therapy (AT) has demonstrated a clear ability to elicit nuanced and personally meaningful expressions for personalization, little work has explored using AT as an approach to generate personalized content for immersive relaxation. For instance, users’ drawings can be transformed into VR environments that reflect their unique emotional landscapes or autobiographical contexts. Additionally, self-reported feelings, emotions, and experiences can be integrated into tailored relaxation guidance, creating a more immersive and personally relevant experience. However, such opportunities have not yet been fully explored.


In this study, we investigated how an art therapy technique called \textit{The Safe Place} can be leveraged to personalize immersive relaxation training. This technique invites users to depict a place that makes them feel safe and relaxed, through both dialogue and sketching. These personal expressions are then used as input to create an AI-assisted, user-authored virtual environment together with personalized audio guidance infused with the user’s life experiences. Specifically, we posed the following two research questions:
    
        \begin{itemize}
            \item RQ1: Quantitatively, whether personalized features derived from an \general{art-therapy-inspired activity} can co-exist with a typical biofeedback-supported mechanism to jointly facilitate users’ relaxation training?
            \item RQ2: Qualitatively, how such new personalization pathways, namely personal life experience-infused audio guidance and a user authored virtual environment, may afford new design opportunities for enriching the relaxation training experience with personally meaningful elements?
        \end{itemize}

To answer these questions, we develop \name, a VR biofeedback relaxation system that leverages an AI-powered \general{art-therapy-inspired} exercise called \textit{The Safe Place}. Users sketch and describe a personal 'safe place', which is then processed through a generative AI pipeline to produce a user-authored 360° virtual environment along with audio guidance infused with their personal life experiences. During relaxation training, users receive breathing biofeedback in their personalized 360° virtual environment, accompanied by personal life experience-infused audio guidance.

For RQ1, we conducted a 2×2 between-subjects factorial study (n=52) using mixed-method. Specifically, we examined four experimental conditions: (1) both personalization features and biofeedback mechanisms present, (2) personalization only, (3) biofeedback only, and (4) neither present. This design enabled us to examine differences in relaxation effects and user experiences across conditions, as well as to assess the main effects of each factor (art-therapy\general{-}derived personalization and biofeedback) and their interaction.

For RQ2, we thematically analyzed the recordings of the interviews and \general{the art-therapy-inspired activity} sessions with participants who experienced relaxation training with art-therapy-derived personalization. This analysis revealed design opportunities, such as integrating verbal and non-verbal inputs to derive complementary and mutually transferable personalization cues, and extracting personally meaningful elements and experiences from autobiographical memory to derive in-depth personalization.
 

The contributions of this work are as follows: 1) We developed \name, a VR-based biofeedback relaxation training system that incorporates user-led in-depth personalization enabled by an AI-infused \general{art-therapy-inspired activity}, including a user-authored 360° virtual environment with personal life experience-infused audio guidance. 2) Through a mixed-method evaluation, we provide an empirical understanding of how such a system influences both the effectiveness of relaxation training and the quality of the user experience. 3) We derived design implications and future directions of utilizing art therapy as a powerful tool for creating highly effective and individualized relaxation experiences, enriching the relaxation training experience with personally meaningful elements.

\section{Background and Related Work}

\subsection{HCI for Relaxation Training}


Relaxation training is a commonly used approach for reducing anxiety, enhancing emotional regulation, and improving overall well-being \cite{manzoni2008relaxation}. To facilitate relaxation training, diverse interactive systems have been developed to provide guidance \cite{1,2,3} and feedback \cite{4,5,6}, supporting users in practices such as breathing regulation \cite{2}, progressive muscle relaxation \cite{3}, body awareness meditation \cite{10}, or mindfulness-based practice \cite{45}. These systems have taken the form of mobile applications \cite{7,9}, web-based platforms \cite{11}, and, more recently, virtual reality (VR) systems \cite{1,3,16,17,18}. Recently, newly-designed relaxation-supporting systems have increasingly emphasized the user experience, with careful consideration on interface design and form-giving. For instance, \textit{Sonic Blankets} \cite{choubey2024sonic} incorporate real-time auditory stimulation modulated by users' movements for promoting relaxation. \textit{GlowGrow} \cite{xue2025glowgrow} used changes in ambient lighting to facilitate deep-breathing–based relaxation for pregnant women. \textit{Embreathe} \cite{haynes2024just} is a huggable cushion that simulates breathing through shape changes, haptically guiding users in mediated breathing to reduce anxiety and provide comfort. These studies highlight that, in relaxation training, user–system interactions serve a dual role: (1) providing guidance and feedback, and (2) fostering relaxing experience. Aligned with this design consideration, VR has become a widely used medium for relaxation applications.


In particular, VR has been widely used as a biofeedback interface in relaxation training due to its ability to immerse users in controlled, multisensory environments \cite{12,16,17} and to couple VR elements to physiological data, supporting biofeedback display and self-regulation. Previous research \cite{23,24} found that VR systems can enhance users' perceived sense of presence, which is positively associated with subjective meditation depth. Moreover, exposure to virtual calming  nature scenes has been shown to effectively increase self-reported relaxation \cite{knaust2022exposure, grassini2022use}. Additional examples can be found in recent HCI studies. For instance, a heart rate variability biofeedback system was developed with a virtual beach scenery at sunset \cite{blum2019heart}. \textit{Stairway to Heaven} offers users a gamified VR journey through a densely forested environment \cite{19}. \textit{Forestlight} immerse users in a realistic rainforest scenery \cite{ zhang2023forestlight}. \textit{DEEP} created virtual game scenes of an underwater world to help reduce anxiety \cite{van2016deep}. While natural environments are among the most commonly used stimuli for relaxation, users’ perceptions and experiences of VR environments can vary widely. Many prior studies \cite{25, heyse2019personalized, xue2025glowgrow} highlight the importance of personalization in relaxation systems. However, few (see \cite{29, 33} as exceptions to be discussed below) have enabled users to self-create virtual environments that feel safe, comfortable, or personally meaningful. \general{More importantly, research in relaxation training and mental health indicates that personalization based on personal experiences and memories can deepen relaxation \cite{Pizzoli_Mazzocco_Triberti_Monzani_AlcañizRaya_Pravettoni_2019}, strengthen emotion-regulation \cite{Miguel-Alvaro_Guillén_Contractor_Crespo_2021}, and foster a greater sense of psychological safety \cite{53}. Despite its potential, such in-depth personalization has rarely been incorporated into biofeedback-supported VR relaxation systems. Therefore, exploring this integration is crucial, as it may allow biofeedback mechanisms to operate within environments that feel inherently safe, resonant, and motivating, thus enhancing the intervention’s impact.  This motivates our research.}

\subsection{Personalization in Relaxation Training}
Early works on personalized user interfaces focused on adapting the interface according to users’ preferences \cite{26}. With the increasing diversity of interaction methods, the concept of personalized interactions has been expanded to multiple dimensions, which may broadly include a wide range of human factors, such as interests, personality, knowledge, culture and past experience \cite{27}.  In the domain of personal informatics and wellbeing, personalized systems have been widely used to support practices such as emotion regulation \cite{28, 29}, and relaxation training \cite{31, 32, 33}. These systems usually provide users with an interface based on a few options or prompts to customize their experience. For example, Collaud et al. \cite{34} developed a multimodal haptic meditation system in which users could personalize their meditation adjusting  parameters with a digital interface or analog controllers and proposed guidelines for designing such interface to support personalization. Halim et al. \cite{29} developed an individualized VR allowing users to personalize avatars, environments, and therapeutic scenarios by an interface, thereby enhancing immersion and fostering self-compassion.

The rise of AI technologies has enabled interventions to be adaptively aligned with users’ preferences and situational contexts, rather than requiring manual configuration through a graphical interface, thereby improving effectiveness and enriching the personalized content \cite{33}. Fang et al. \cite{30} introduced a system that generates emotionally expressive responses (e.g., reflection, encouragement) in a user’s own cloned voice, through which users could vividly envision improved versions of themselves, boosting emotional states, resilience, and motivation while fostering a stronger sense of self-compassion and goal commitment. For relaxation training, Kim et al. \cite{39} developed a system generating the contents of meditation using an LLM to generate meditation scripts based on Acceptance and Commitment Therapy (ACT) and then turn these scripts into spoken words, simulating a "self-talk" technique. Nguyen et al. \cite{33} developed an AI-powered meditation system. With provided context, meditation type, general mood or goals for the meditation, their system provides personalized immersive experiences that integrate text, audio, and visuals. However, the perceived level of personalization was limited, as the system offered only a small set of contextual options during their study. These works highlight the potential of LLM-based personalization, which operates semantically and can process non-verbal input without relying on rigid rule-based algorithms. Yet, little work has explored how LLMs might fuse autobiographical memories and verbalized imagination into tailored environments and personalized audio guidance, which is an opportunity we seek to investigate in this study.

Customizing the virtual environment is a common personalization approach in immersive wellbeing technologies \cite{29, 25, 31, Wagener_Niess_Rogers_Schöning_2022}. To enrich personalization options, some studies allow users to participate in creating or modifying the virtual environment. Cheung et al. \cite{36} explored how people living with dementia could co-create meaningful environments through collaborative art-making of 2D collages and model boxes, which were then transformed by the researchers into abstract virtual reality spaces that fostered emotional engagement and well-being. Similarly, Ahn et al. \cite{35} designed OWN, a personalized virtual space for introspection, created through a co-design process in which users shared personal memories and selected meaningful objects, sounds, and places that were then incorporated into the virtual environment, fostering self-reflection and emotional well-being. A designer and two psychologists participated in the co-design, and a programmer also contributed to building the virtual space. These studies explored user-engaged personalization, where users are not merely operating an interface or selecting from a dataset, but can adapt the virtual environment to their personal memories, emotions, and lived experiences. The in-depth personalization related to the users themselves can further foster the effects and engagement in activities that supports mental health. However, such in-depth personalized systems often require a significant amount of time (multiple interventions may take several weeks \cite{36}) and the contributions and labors by professionals from multiple disciplines, which could be resource-intensive for relaxation training. With the advancements of generative AI (Gen-AI) technologies, new opportunities have emerged to lower the barriers to image creation. An increasing number of studies \cite{436, 437} have explored AI-assisted, user-authored imagery in the context of mental wellbeing. However, little work to date has examined how such approaches can support user-engaged personalization. Our study aims to investigate whether, and in what ways, AI-assisted user-authored VR environments can be meaningfully applied in this domain.


\subsection{\general{Safe Place Technique and Art Therapy in HCI}}

\general{Art therapy is based on the concept that the act of creating art fosters healing and recovery, acting as a powerful means of expressing thoughts and emotions without words \cite{201}. Classic art therapy literature highlights how drawing, painting, collage, and multimaterial making serve as therapeutic "languages' that help individuals process thoughts and feelings that may be difficult to articulate verbally \cite{Rubin_2012}. Art therapy is not limited to the making of final artistic product; rather, the creative process, including exploration, transformation, and reflection, is central to therapeutic change. This view is formalized in the Expressive Therapies Continuum (ETC), which conceptualizes how clients engage with materials across sensory, affective, perceptual, and cognitive–symbolic levels \cite{Hinz_2019}, and is widely used in both individual and family-based expressive arts therapy \cite{Malchiodi_2005}.}

\general{Within the broader landscape, guided imagery and visualization represent important modalities in art therapy \cite{206, Dolce1999TheUO, Barnett_Vasiu_2024}. These approaches invite individuals to construct internal scenes that evoke safety and calmness \cite{utay2006guided, hart2008guided, krau2020multiple}. One of the most established forms is the Safe Place (or Calm Place) technique, which guides individuals to imagine and elaborate a personally meaningful environment associated with comfort and security \cite{Zehetmair_Nagy_Leetz_Cranz_Kindermann_Reddemann_Nikendei_2020, Drujan_Fallgatter_Batra_Fuhr_2023}. Safe-place imagery has long-standing applications across art therapy \cite{53}, emotion regulation \cite{Drujan_Fallgatter_Batra_Fuhr_2023}, and stress-reduction practices \cite{Bonadies}. The process typically begins by guiding the individual into a relaxed state, often through deep breathing. In the first stage, the therapist leads the client to imagine a safe and comfortable environment drawn from their past experiences and memories. In the second stage, the therapist encourages the client to continue imagining and mentally engaging with various sensory details of the safe place, including visual, auditory, olfactory, gustatory, and tactile sensations. This phase helps the individual experience a more immersive connection to the imagery. The final stage involves using art-making tools to visually represent the safe place, externalizing the positive emotions and memories associated with it \cite{Rappaport_2009, 201}. Research shows that constructing a personalized imagery safe place  by drawing from autobiographical experiences \cite{Boccaccio_Pennisi_Guerrera_Platania_Torre_Varrasi_Vezzosi_Coco_Castellano_Pirrone_2024} , sensory memories \cite{Holmes_Mathews_2010,Boccaccio_Pennisi_Guerrera_Platania_Torre_Varrasi_Vezzosi_Coco_Castellano_Pirrone_2024}, or meaningful landscapes \cite{Bosgraaf_Spreen_Pattiselanno_Hooren_2020,53} can enhance the effectiveness of reducing stress responses \cite{Bosgraaf_Spreen_Pattiselanno_Hooren_2020, 53}, regulate emotion \cite{Bosgraaf_Spreen_Pattiselanno_Hooren_2020, Holmes_Mathews_2010} and strengthening psychological safety \cite{53} during therapeutic work.}

\general{Recent HCI research has increasingly explored how digital and generative technologies can extend art therapy practices, from early efforts showing how digital media enhance accessibility and creative engagement \cite{Collie_Čubranić_2002, DarewychOlena2018DTUi} to more recent studies exploring AI-assisted image generation and co-creative drawing as expressive materials. Systems such as \textit{DeepThInk} \cite{436} show how generative AI can lower the threshold for art-making, expand expressive repertoires, and support synchronous or asynchronous digital art therapy. Similarly, family expressive-arts-therapy systems blend traditional physical materials with generative AI to help children and parents externalize emotions through storymaking and symbolic imagery \cite{402}. Complementing these, \textit{TherAIssist} \cite{Liu_Bai_Zhang_Zhang_Zhang_Zhao_An_2025} explores how AI can support between-session art therapy homework, combining co-creative image generation with conversational agents to help clients articulate emotions and enabling therapists to customize therapeutic prompts and review AI-compiled records. These works highlight a growing opportunity for HCI systems to leverage AI not only as an art generator, but also as a novel expressive material and therapeutic scaffold, supporting emotional articulation and symbolic exploration.}

\general{Despite these advances, most existing HCI research has focused on how digital or AI-based tools can facilitate the art therapy activities themselves. However, little work has explored how art therapy  activities and mechanism might support personalizing relaxation training in non-clinical contexts. In particular, while the art therapy and guided imagery literature emphasizes the therapeutic power of autobiographical experiences, sensory memories, and personally meaningful landscapes, these elements have rarely been leveraged to shape personalized relaxation training. Given the demonstrated significance of these elements in stress reduction and emotion regulation, we therefore investigate this underexplored pathway. Our work takes an early step toward integrating art-therapy-derived personalization rooted in users’ own safe-place imagery with immersive biofeedback to support more resonant, meaningful, and potentially more effective relaxation experiences.}

\section{\name}

\subsection{System Design}


In this study, we present \name, a VR–based biofeedback relaxation system that incorporates user-led in-depth personalization enabled by an AI-infused \general{art-therapy-inspired activity}. This personalization process generates an AI-assisted, user-authored 360° virtual environment aligned with the user’s imagined sanctuary, along with guidance audio enriched by the user’s life experiences as well as emotional and sensory memories. As shown in Fig.\ref{fig:teaser}, relaxation training with \name{} consists of two phases: personalization and relaxation. During the personalization phase, users engage in a Safe Place practice that generates a tailored 360° environment map using generative AI (gen-AI) tools, based on both verbal and non-verbal descriptions. In parallel, large language models (LLMs) process users’ verbalization about their Safe Place, together with associated personal life experiences and emotional or sensory memories to produce personalization inputs. These inputs are then used to generate personalized audio relaxation guidance infused with users’ life memories. During the relaxation phase, users are immersed in their self-created 360° virtual environment, accompanied by the personalized audio guidance, and perform breathing-based relaxation training with particle effects that respond to their breath as biofeedback.




\begin{figure*}[h]
  \centering
  \includegraphics[width=\linewidth]{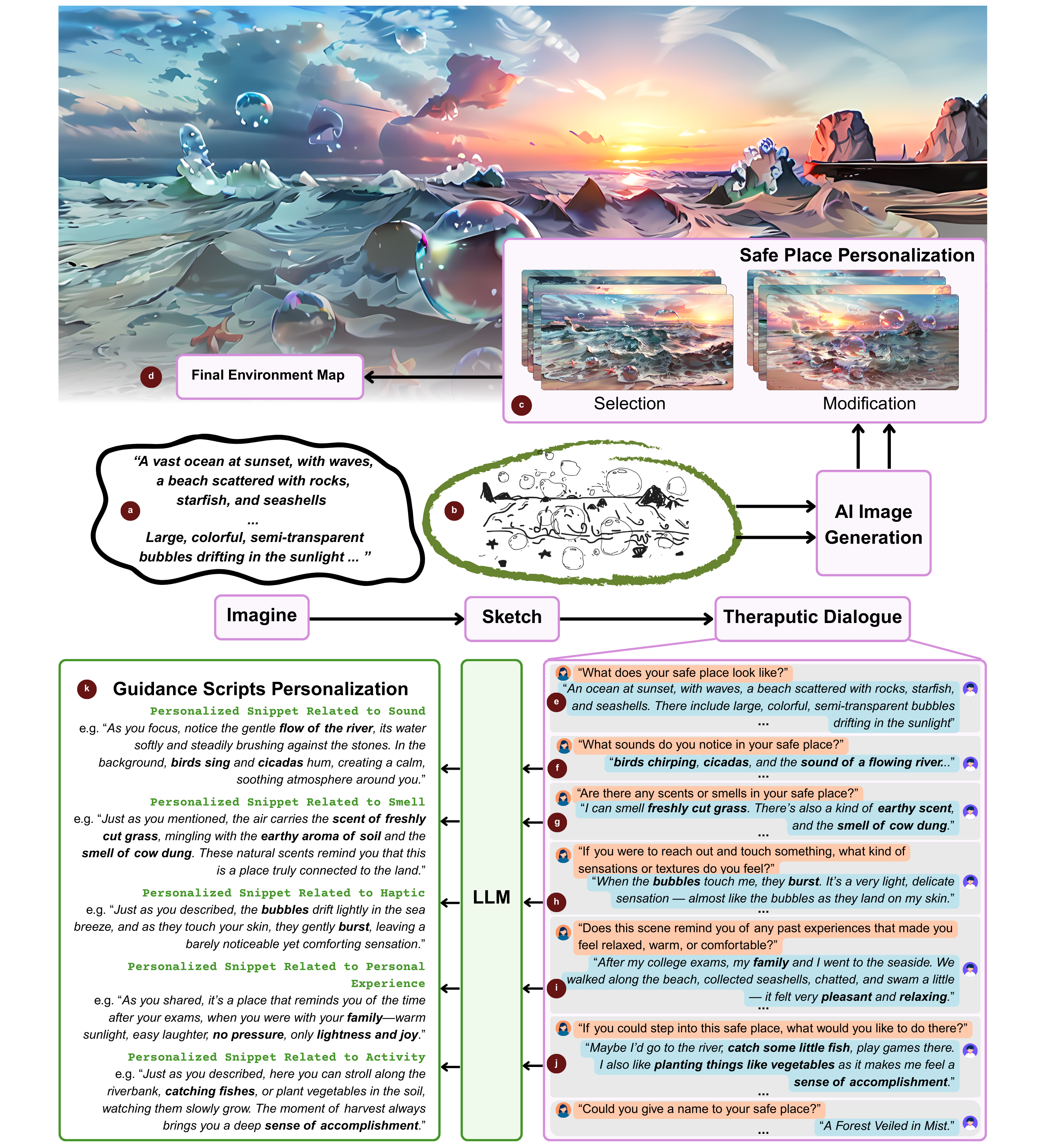}
  \caption{The personalization phase of \name{} system.}
  \Description[A diagram of the personalization phase]{In the middle, the user first describes their imagined safe place using key elements (Figure 2(a); e.g., an ocean at sunset with waves, a beach with rocks and shells, and bubbles), and then represents these elements with simple sketches (Figure 2(b)). The user’s therapeutic dialogue with the art therapist is also recorded, shown as multiple dialogue snippets from different users in Figure 2(e–j) at the bottom. These dialogue snippets are input into an LLM to generate corresponding personalized guidance scripts, shown in Figure 2(k). For example, a dialogue about sound (Figure 2(f)) produces a guidance script focused on sound, while a dialogue about personal experience (Figure 2(i)) produces a guidance script focused on personal memories and feelings. At the top, Figure 2(c) shows how the user co-creates a personalized 360° environment map with an AI image generation tool by first selecting one image from a set of text-generated options, then editing the chosen image to adjust visual elements. After these modifications, the user obtains the final personalized environment map shown in Figure 2(d).}
  \label{fig:SystemDesign}
\end{figure*}

\subsubsection{Personalization Phase}


In the personalization phase, the user attends an \general{art-therapy-inspired activity} session based on the Safe Place technique \cite{206} , which could help users imagine and visualize a personal sanctuary, enabling them to externalize the imagery, experience, sensations and memories that make them feel safe, relaxed and comfortable, from which in-depth personalization inputs can be extracted. The personalization phase proceeds as follows:

\textbf{Imagine:} To start, the art therapist guides the user to imagine a place that makes them feel safe and relaxed with a typical Safe Place guidance script. \general{This script is translated and adapted from the clinical guided-imagery script\footnote{Sicherer Ort\_02.10.2020\_md.docx. https://zenodo.org/records/7415323} used in a previous study \cite{Drujan_Fallgatter_Batra_Fuhr_2023}.} Under the guidance, The user has 5 minutes to imagine their 'safe place' and any details related to it such as weather, objects, sensations and events (e.g., see Fig.\ref{fig:SystemDesign} (a)).

\textbf{Sketch:} The user is then asked to sketch their ‘safe place’ on smart devices or on paper and briefly describe the key visual elements in their sketches that make them feel relaxed (see F.\ref{fig:SystemDesign} (b)). \general{The user-created sketch is digitized (scanned or photographed) and used as input reference to produce a corresponding 360° environment map reflecting the participant’s safe-place imagery.}

\textbf{Therapeutic Dialogue:} During environment map generation, the therapist will have a therapeutic dialogue with the users, inviting them to elaborate on the imagery (see Fig.\ref{fig:SystemDesign} (e)) and senses they’ve imagined, exploring what they can hear (see Fig.\ref{fig:SystemDesign} (f)), smell (see Fig.\ref{fig:SystemDesign} (g)), touch (see Fig.\ref{fig:SystemDesign} (h)), and even taste. The user will further detail why they have imagined and visualized such their 'safe place', their past experience about it (see Fig.\ref{fig:SystemDesign} (i)) as well as the activities they would like to do in their 'safe place' (see Fig.\ref{fig:SystemDesign} (i)). Finally, the user gives a name to their safe place. 

\textbf{AI-Assisted \general{Co-Creation:}} \general{Concurrent with the therapeutic dialogue, a researcher inputs the user’s sketch (as reference) and verbal description (as prompts) into an AI-based image generation tool to generate 360° environment maps representing their ‘safe place’. The user then engages in an active co-creation process. As shown in Fig.\ref{fig:SystemDesign} (c), at first the user can select the one they prefer most from the candidates. After making a selection, the user is allowed to modify the environment through the AI image generation tool with the help of the researcher. They can correct misgenerated image elements (e.g., replacing waves with rocks), add or remove elements, and adjust global attributes such as weather, time of day, sunlight intensity, or wave dynamics. Through this iterative co-creation process, the resulting 360° environment incorporates the key imagery described by the users, such as sunsets, beaches, or floating bubbles (see Fig.\ref{fig:SystemDesign} (d)). Once the user is satisfied with the resulting 'safe-place' , it will be projected to the VR environment for the relaxation phase.}

\textbf{AI-Enabled Audio Guidance Personalization:} The \general{art-therapy-inspired activity} session is recorded. After the session, recorded and transcribed therapeutic dialogue texts are processed by an LLM to generate a set of personalized relaxation scripts that resonate with users’ personal experiences and sensory cues of relaxation. For example, Fig.\ref{fig:SystemDesign} (k) illustrates different personalized snippets generated by the LLM, each corresponding to the therapeutic dialogues from various scenes and contexts (see Fig.\ref{fig:SystemDesign} (e–j)). These scripts are further merged with an evidence-based relaxation guidance template (see Appendix A), which is then transformed into audio guidance using a text-to-speech (TTS) model.

\subsubsection{Relaxation Phase}
\begin{figure*}[h]
  \centering
  \includegraphics[width=\linewidth]{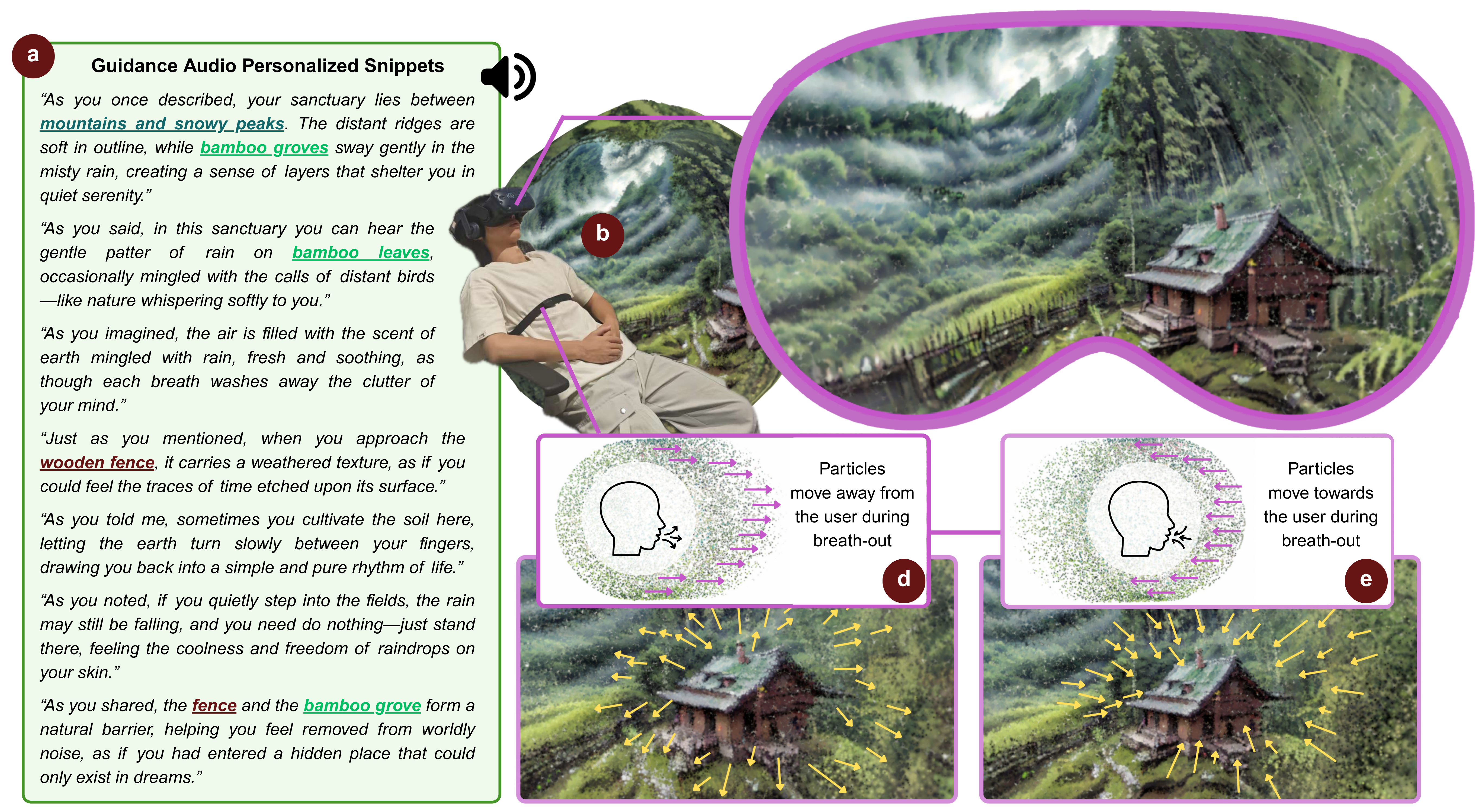}
  \caption{The relaxation phase of \name{} system: (a) Personalized audio guidance with embedded snippets is played; (b) the user wears an HMD and a breath belt to immerse in the virtual environment; (c) the environment is presented with particle effects; (d) during inhalation, particles move toward the user; and (e) during exhalation, particles move away.}
  \Description[A diagram of the relaxation phase]{On the left, Figure 3(a) shows text examples of personalized audio guidance with embedded snippets. The snippets explicitly mention key elements of the user’s safe place (e.g., bamboo groves and wooden fences), highlighted in colored text to indicate where personalized details are inserted. On the top right, Figure 3(b) illustrates the user wearing a head-mounted display (HMD) and a breathing belt while immersed in the virtual environment. The environment is represented as a spherical space made of particles, with the user’s 360° environment map displayed on the inside surface, and a first-person view preview is shown next to the illustration. On the bottom right, two biofeedback illustrations show how breathing affects the particle motion: Figure 3(d) depicts inhalation, where particles move toward the user, and Figure 3(e) depicts exhalation, where particles move away from the user. Each biofeedback illustration is paired with a corresponding first-person preview showing the same inward or outward particle movement.}
  \label{fig:Relaxation}
\end{figure*}

Before the relaxation training begins, the user sits in an armchair, wears a breathing belt sensor, and puts on a VR head-mounted device (HMD). The panoramic image is rendered on a large sphere composed of floating, moving particles that surround the user and visualize their real-time breathing. (see Fig.\ref{fig:Relaxation} (b)).  \general{We grounded our visualization choices in prior studies that showed benefits of integrating physiological feedback into the environment itself rather than displaying it on separate charts or panels \cite{blum2019heart, rockstroh2021mobile}.  Some biofeedback systems \cite{16, Prpa_Tatar_Françoise_Riecke_Schiphorst_Pasquier_2018} visualize user's biofeedback data internal with ambient properties of the virtual or mixed environment, showing that such environment-based biofeedback displays support mindfulness, interoceptive awareness, and sustained engagement by letting users “feel” biofeedback through the world around them instead of monitoring separate UI elements.}

\general{\textbf{During the relaxation training, users are immersed in a panoramic visualization of their 'safe place', within which real-time breathing biofeedback is displayed.} The user's breathing pattern is visualized through the natural movement of particles embedded in the 360° scene (see Fig.\ref{fig:Relaxation} (c)). Specifically, the floating particles move towards the user's body during breath-in (see Fig.\ref{fig:Relaxation} (d)), and spread outward during breath-out (see Fig.\ref{fig:Relaxation} (e)). The use of particle motion to represent users’ breathing was inspired by prior work on \textit{DeLight} \cite{yu2018delight}, which suggests that such natural mapping is intuitive to understand and effective in facilitating breathing regulation.}



\textbf{Alongside the breathing biofeedback, the user listens to their personalized relaxation guidance:} in addition to general deep breathing guidance, the personal audio guidance includes personalized snippets corresponding to the visual elements in their 'safe place' to help the user become aware of and adjust their breathing. For example, in Fig.\ref{fig:Relaxation} (a), the personalized snippets repeatedly emphasized the imagery of the fence and bamboo grove, as these elements were identified as crucial for the user to feel relaxed and safe. The personalized audio guidance not only highlighted the presence of these elements but also guided the user to perceive the fence and bamboo grove as forming a natural barrier that protects them from interruptions. Moreover, these snippets were structured to follow the sequence of \textit{The Safe Place} technique, helping the user revisit familiar sensations and activities as if they were entering this ‘safe place’ again.




\subsection{System Implementation}

\subsubsection{Personalization of VR-based Relaxation Training}
\general{Two professional art therapists (T1 and T2; both self-identified female; ages: T1 = 24, T2 = 49) participated to support our study. T1 has 2 years of experience and over 100 prior cases, and T2 has 8 years of experience with more than 350 cases. Both hold a Master’s degree in Art Therapy. T1 is a member of our research team, and T2 is a former collaborator. T1 designed the materials for the art-therapy-inspired activity, including the protocols, the adapted Safe Place guidance script, and the open-ended therapeutic dialogue script. T2 subsequently reviewed these materials to ensure their professional accuracy and therapeutic appropriateness. The final scripts and protocols are provided in the supplementary material.}

The user's transcribed therapeutic dialogue texts are fed into the OpenAI GPT-4o for generating relaxation guidance scripts. In the predefined prompt, the chatbot is asked to first extract the key information about the user's safe place, then to imitate and replace the selected snippets (usually 6 to 8 sentences) in a given meditation guidance script template referenced from \cite{321} (see Appendix A) with sentences with the same style. (The full prompt is provided in Appendix B). \general{The generated script is moderated by an art therapist to ensure safety and appropriateness before use, and minimally adjusted only when necessary.} The \general{final} script is then synthesized into audio using GPT-SoVITS \cite{322}, fine-tuned with a 5-minute dataset sliced from UCLA Health’s guided meditation audio (English) \cite{323}.


We help the user create personalized virtual environment through Stable Diffusion web UI, which is a web interface for text to image generation. The base model we use is \textit{Inazuma V10}\footnote{https://civitai.com/models/76988?modelVersionId=81783}, a Stable Diffusion v1-5 (SD 1.5) fine tune (also known as checkpoint), which allows us to generate image in semi-realistic style. To guide the diffusion model to generate 360 panorama environment map, we use a Latent Labs 360 LoRA \footnote{https://civitai.com/models/10753/latentlabs360}. In the personalization phase, the keyword description of the visual elements in the user's safe place is input as text prompts. And the user's sketch is input to a ControlNet to guide the visual structure, making the generated environment map closer to the user's imagination.  During the \general{art-therapy-inspired activity} session, we generate the environment maps using a Latent Consistency Model (LCM) LoRA \footnote{https://huggingface.co/papers/2311.05556} and the corresponding sampler with a preview parameter preset: (Size: 1024x512, Steps: 20, Sampler: LCM, CFG Scale: 7). This ensures that the user can immediately preview the generated map after each revision. A final version is generated after the \general{art-therapy-inspired activity} session with a high-quality parameter preset: (Size: 2048x1024, Steps: 30, Sampler: DPM++ 2M Karras, CFG Scale: 7).

\subsubsection{ASafePlace Relaxation Training Application}

The \name{} VR application was developed using Unity's Universal Render Pipeline (URP). It runs on a desktop PC with an VR HMD (HTC Vive Pro Eye). Visual Effect Graph (VFX Graph) in Unity was used to create the surrounding sphere which present personalized environment map. Specifically, we first input a sphere mesh to shape the particle effect and then sampled the color from the input environment map for each particle at different positions of the sphere. As a result, the user can see the panorama view displayed by particle effects. For the biofeedback, we passed the real-time breathing data captured by the breath belt into the VFX Graph to control the movement of the particles. The real-time breathing data value is mapped to the local velocity of each particle. The breath sensor belt (Elastech\general{\footnote{http://www.elas-tech.com}}) communicate with the applciation via Bluetooth connection. The audio relaxation guidance is played separately using the deafault audio player of Windows 11.





\section{User Study}

In the user study, we used a mixed-method approach to answer our research questions RQ1 and RQ2. Quantitatively, we collected participants’ physiological and subjective relaxation metrics to examine whether personalized features derived from \general{art-therapy-inspired activities} can complement a standard biofeedback mechanism to jointly enhance relaxation training (RQ1). Qualitatively, we use semi-structured interview to gain insights into how such personalization via \general{art-therapy-inspired activities} may afford new design opportunities for enriching the relaxation training experience with personally meaningful elements (RQ2).

 Specifically, we conducted a 2×2 between-subjects factorial experiment, with personalization via \general{art-therapy-inspired activities} (P) and biofeedback (B) as the two independent variables. The participants were randomly assigned into 4 groups as shown in Table \ref{tab:conditions}.  In Group PB (Personalization + Biofeedback), participants created a personalized immersive VR environment and relaxation guidance \general{audio} via an \general{art-therapy-inspired activity} session and then completed a relaxation session with breathing biofeedback. In Group PN (Personalization + No Biofeedback), participants also completed the personalization phase and relaxation phase, without receiving breathing biofeedback. In Group NB (No Personalization + Biofeedback), participants skipped the personalization phase and instead selected a pre-generated 360\general{°} environment for relaxation phase, during which they also received breathing biofeedback. In Group NN (No Personalization + No Biofeedback), participants only complete relaxation session with a selected a pre-generated 360\general{°} environment. They did not receive breathing biofeedback. \general{At the same time, the participants in Groups NB and NN were guided by a standard template audio (see Appendix A). This ensured that the audiovisual stimuli used for relaxation training were comparable across groups while eliminating the personal relevance and connection of the stimuli in Groups NB and NN. All groups, including NN, used a VR environment to ensure comparable conditions. }


\begin{table*}[h]
  \centering
  \caption{The conditions in our 2×2 factorial design.}
  \label{tab:conditions}
  \Description[A table illustrates the four conditions.]{Dimension 1: No Personalization/ With Personalization. Dimension 2: (No biofeedback/ With Biofeedback). Four cells representing the four conditions: No Personalization + No Biofeedback (NN), Personalization + No Biofeedback (PN), No Personalization + Biofeedback (NB), Personalization + Biofeedback (PB)"}
  \begin{tabular}{p{3.3cm} p{3.3cm} p{3.3cm}}
    \toprule
    Factor Level & No Biofeedback & With Biofeedback \\
    \midrule
    No Personalization  & 
      \makecell[l]{No Personalization  + \\ No Biofeedback \\\quad\quad(NN)} & 
      \makecell[l]{No Personalization  + \\ Biofeedback \\\quad\quad(NB)} \\
    With Personalization & 
      \makecell[l]{Personalization  + \\ No Biofeedback \\\quad\quad(PN)} & 
      \makecell[l]{Personalization + \\ Biofeedback \\\quad\quad(PB)} \\
    \bottomrule
  \end{tabular}
\end{table*}

\subsection{Participant and Procedure}
\general {We recruited 52 participants (N=52) through social networks (15 male, 37 female; aged between 18 and 42; mean=23.67, SD=5.29).  All participants were from a non-clinical population. Three participants reported prior experience with VR devices, and two had previous exposure to art-therapy or meditation practices. Detailed demographic information (e.g., gender and age distribution, condition allocation, and participant IDs per group) is available in the supplementary material. At the start of each session, the art therapist conducted a brief mental-health check to ensure that the participant did not present vulnerabilities that could pose risks during the art-therapy session.}

\begin{figure*}[h]
  \centering
  \includegraphics[width=\linewidth]{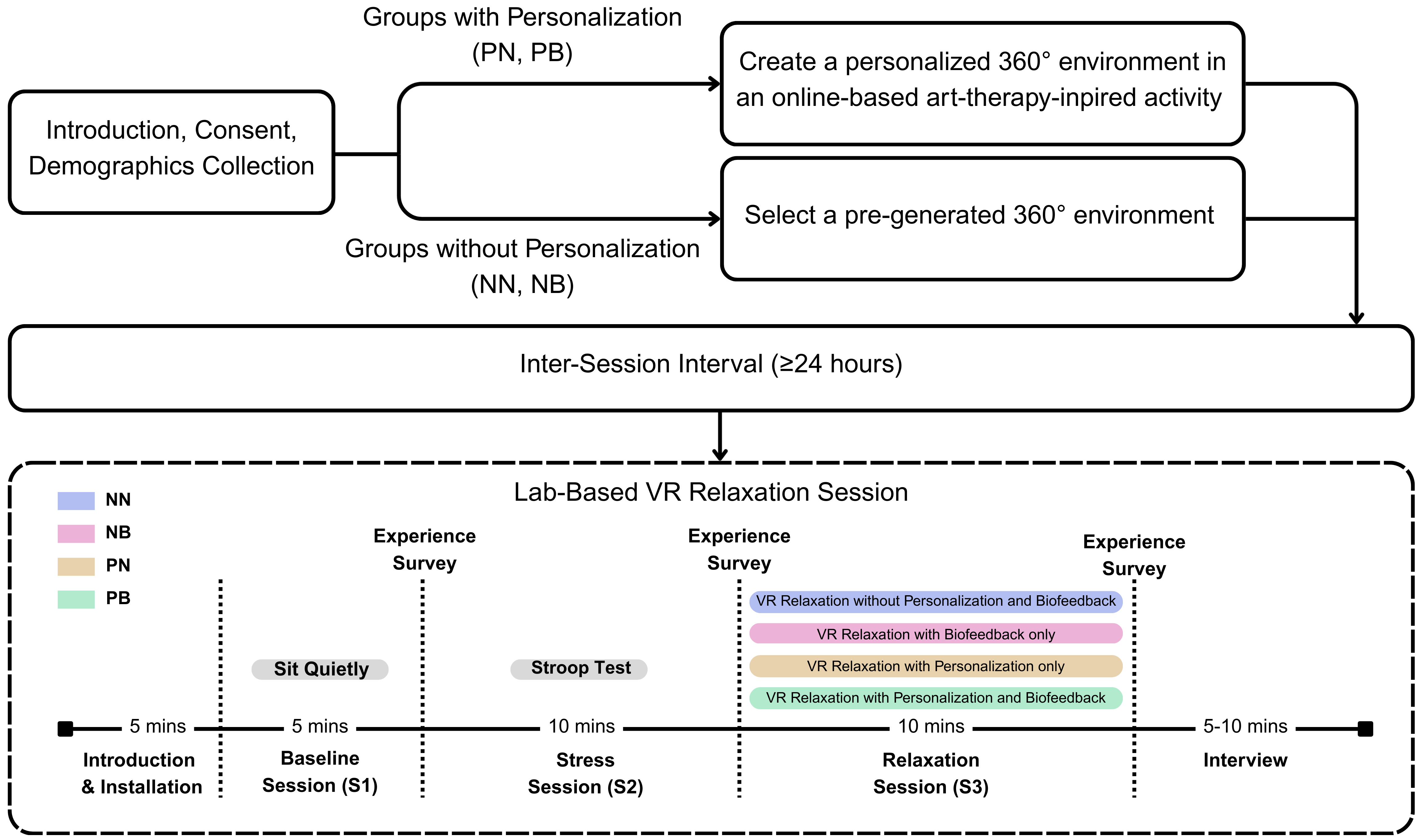}
  \caption{The procedure of the user study.}
  \Description[A flow chart of the user study with two connected parts.]{An art-therapy-inspired activity session (top) and a lab-based VR relaxation session (bottom). In the top procedure, participants begin with an introduction, then provide informed consent and complete demographic questions. Next, they are randomly assigned to one of four experimental conditions. Participants in the two personalization conditions (PN and PB) create a personalized virtual environment through an online art-therapy-inspired activity, while participants in the two non-personalization conditions (NN and NB) select a pre-generated virtual environment. A horizontal bar in the middle of the chart indicates that all participants have an interval of more than 24 hours between the two sessions. The bottom procedure shows the lab-based VR relaxation session, which includes an introduction and equipment installation, followed by three sequential sessions: a baseline session (s1), a stress session (s2), and a relaxation session (s3), and then a final interview. In s1, participants sit quietly for 5 minutes. In s2, participants complete a 10-minute Stroop test to induce stress. In s3, participants complete a 10-minute relaxation training session in VR based on their assigned condition.}
  \label{fig:StudyProcedure}
\end{figure*}

Fig.\ref{fig:StudyProcedure} shows the procedure of the study. \general{A researcher first contacted participants through social networks, provided the study briefing and privacy policy, and directed them to an online consent form that included demographic questions. After registration, participants were randomly assigned to one of the four conditions and scheduled their experimental sessions. Each participant in the groups with personalization (PN and PB) created their personalized 360° environment maps in an online art-therapy-inspired activity session. Each participant in groups with personalization (NN and NB) selected a pre-generated 360° environment map (from the ones created during the personalization phase of Groups PB and PN). To ensure consistency across conditions and to avoid potential short-term influence of the art-therapy-inspired activity, all groups had a $\geq$24 hour interval before their lab-based VR relaxation session.} 

\general{In the lab-based VR relaxation session,} all conditions shared the same initial procedure consists of three sessions: baseline session (S1), stress session (S2) and a relaxation session (S3).  The experiment began with a 5-minutes instruction and installation session for participants to learn what they will do and wear the wearable devices. In S1, the participants sat quietly in a 5-minutes baseline session to help researchers to get the baseline physiological data (heart rate variability and respiration rate). This is followed by a stress session (S2) during which they were asked to complete the Stroop color-word test \cite{stroop} for 10 minutes to induce stress responses. \general{The Stroop color–word test is a validated and widely used cognitive-stress induction task in prior research \cite{3, Lin_Kunze_Ueki_Inakage_2020, Tan_Schöning_Luyten_Coninx_2014, blum2019heart}. Given the goal of our experiment, the Stroop test provides a controlled and reproducible way to examine the effectiveness of different relaxation conditions.} In S3, participants completed a 10-minutes relaxation training session \general{according to their assigned experimental condition}.

After the baseline, stress, and relaxation sessions, participants completed an experience survey consisting of several subjective questionnaires to investigate user's subjective level of relaxation, stress, sense of flow, presence and personal relevance. \general{Prior work on relaxation training commonly uses three-session protocols with session durations ranging from 5–10 minutes \cite{3, 47, zhang2023forestlight, Alvarsson_Wiens_Nilsson_2010}. Following this established practice, we adopted a 10-minute stress session and a 10-minute relaxation session to maintain consistency and comparability with the existing literature.} Meanwhile, there was a semi-structured interview for each participant to collect their feedback on the relaxation training. All participants were compensated with a locally equivalent amount of 20 USD for their participation. 

\subsection{Set-up}


 \general{The participants in all groups completed the relaxation phase in a dedicated room. All participants sit on an armchair and wore a VR headset during relaxation. Since prior work has already established immersive VR as an effective medium for supporting relaxation \cite{23, 24, grassini2022use, knaust2022exposure, Pancini_Di} and biofeedback engagement \cite{17, 16, 1, 3, zhang2023forestlight, blum2019heart, rockstroh2021mobile, Prpa_Tatar_Françoise_Riecke_Schiphorst_Pasquier_2018}, our study focuses on the impact of personalized features on the relaxation effects within biofeedback-supported VR relaxation training rather than reassessing the effectiveness of biofeedback or VR themselves. Therefore, this setup was kept consistent across all conditions to control for the effects of the VR medium and allow us to isolate the influence of personalized elements and biofeedback mechanisms on the immersive relaxation experience.}

\subsection{Measurements}
\subsubsection{Quantitative data collection}

\general{Following prior studies \cite{3, 47, 49}, we choose the following quantitative measurements.} We collected participants’ physiological data, including heart rate variability (HRV) and respiration rate, across all three sessions (baseline, stress, relaxation) to assess the effectiveness of \name{} on relaxation training. \general{The respiration rate is recorded using a breath belt stretch sensor from \textit{Elastech}\footnote{Elastech Breathing Belt sensor: https://en.elas-tech.com}. This breathing sensor was also used for real-time breathing feedback during the relaxation session. HRV was recorded with a noninvasive photoplethysmography (PPG) sensor\footnote{Pulse Sensor, https://pulsesensor.com/} placed on the participant’s left index finger \cite{Lu_Yang_Taylor_Stein_2009}.} In this study, we use SDNN (Standard Deviation of NN intervals) as the primary HRV index, given its effectiveness in reflecting autonomic balance and relaxation \cite{kim2018stress}. Subjective relaxation and anxiety were measured using the Relaxation Rating Scale (RRS) \cite{RRS} and the State-Trait Anxiety Inventory—State form (STAI-S) \cite{STAI}. To capture participants’ experience during the VR sessions, we administered the Flow State Scale (FSS) \cite{FSS} for flow assessment and the IGroup Presence Questionnaire (IPQ) \cite{IPQ} to evaluate the sense of presence within the AI-generated virtual environments. In addition, the user's perceived relevance was measured by a questionnaire consisting of four items adapted from prior work \cite{Relevance}: (1) The scene in the VR relaxation session was relevant for me; (2) The meditation scripts in the VR relaxation session were relevant for me; (3) The meditation scripts in the VR relaxation session grasped my attention; and (4) The meditation scripts in the VR relaxation session said something important to me. Participants rated each item on a 7-point Likert scale, with higher values indicating greater perceived personal relevance of the VR environment and audio guidance.


\subsubsection{Qualitative Data Gathering}
We conducted a semi-structured interview with an average time of five minutes at the end of the relaxation training. Participants were asked whether and how the training made them feel relaxed, their preferred and least preferred experiences, impressive elements, impressions of the visual effects and audio guidance, and any other feedback they wished to share.  In addition, for participants in the personalized conditions (PB and PN groups), we revisited the earlier \general{art-therapy-inspired activity} sessions. This included reviewing the therapeutic dialogue transcripts and personalized guidance scripts as well as the sketches and AI-generated environment maps they created during the Safe Place exercise. These interviews and art therapy artifacts provided a multifaceted perspective on participants’ experiences, offering both reflective accounts of the relaxation training and personalized representations of their imagined safe places.

\subsubsection{Data Analysis}
For quantitative measurements, we first assessed the normality of the data for each variable using the Shapiro-Wilk test. Within each group, we compared the measures between the stress and \general{relaxation} sessions using a paired t-test (when normality was confirmed) or a Wilcoxon test (for non-normal data). Then, for each group, we calculated the percent changes of the measures in the \general{relaxation} session (relative to their levels during the stress session) and used ANOVA to test which of the conditions worked the best and to test whether the differences between the groups are significant. Finally, a two-way ANOVA (General Linear Model/ univariate analysis in SPSS) was conducted to compare the main eﬀects of \general{art-therapy-derived personalization} and biofeedback and the interaction eﬀect between \general{art-therapy-derived personalization} and biofeedback on the \general{relaxation} practice.

For qualitative data, we verbatim transcribed the recordings of the interview and \general{the art-therapy-inspired activity} session into texts in Google Docs. We used thematic analysis \cite{438} to analyze the data. One author reviewed and coded each participant's interview recording. The \general{art-therapy-inspired activity} session recordings were coded by two author, and another author joined to discuss and organize the final codes with the interview codes, which resulted in the themes presented in \S 5.2.

\subsection{Ethical Considerations}
This study received approval from our institution’s ethics review board. All participants were from a non-clinical population and provided informed consent prior to participation. To safeguard psychological well-being during the art-therapy–inspired activity session, the art therapist conducted a brief mental-health check-in at the start of each session. Participants could pause or withdraw at any time, and a distress-response protocol was in place, though no such cases occurred.

Because the personalization process involved sensitive personal content, all dialogue and sketches were anonymized before analysis or use in the AI-assisted co-creation pipeline. Only de-identified textual descriptions were transmitted to OpenAI GPT-4o; no raw audio, identifying information, or physiological data were shared. The AI tool was used in compliance with its privacy policy and did not store study data. All research data were stored on a local, access-controlled drive, and Google Forms/Docs were used only for consent and demographics.

Given the involvement of therapeutic dialogue and personal imagery, we acknowledge the sensitivity of the data and discuss this in \S 6.4. We outline mitigation strategies and emphasize that our system is intended for research with non-clinical populations rather than clinical therapeutic deployment.

\section{Results}

\subsection{Quantitative results}

\subsubsection{HRV-SDNN}
\begin{figure}[h]
  \centering
  \includegraphics[width=\linewidth]{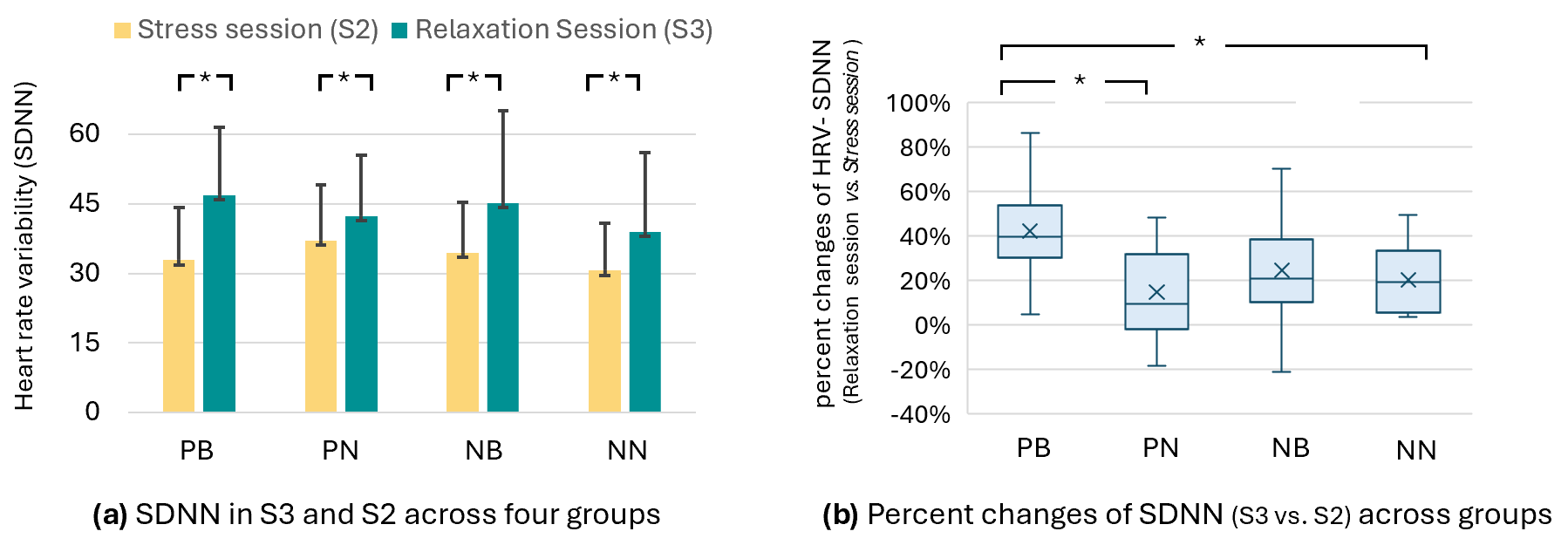}
  \caption{(a). HRV-SDNN changes (Relaxation session S3 \textit{vs.} Stress  session S2) across four groups. (b) Boxplots of percent changes in SDNN between S3 and S2}
  \Description[Heart rate variability during stress and relaxation session across conditions.]{Two plots compare four groups (PB, PN, NB, NN). The left panel is a bar chart showing that heart rate variability is higher in the relaxation session (S3) than in the stress session (S2) for all four groups. A significant increase in SDNN was observed across the four groups. The right panel is a box plot showing the percent change from S2 to S3 for each group, where the personalization-with-biofeedback group shows the largest increase compared to the other conditions. In PB group, SDNN increased significantly larger than PN and NN groups.}
  \label{fig:SDNN_results}
\end{figure}

As shown in Fig.\ref{fig:SDNN_results}(a), the heart rate variability (SDNN) increased significantly during the relaxation session (S3) compared to the stress session (S2) in all groups (PB: 42.5± 20.1\%; PN: 15.0± 21.4\%; NB: 24.7± 24.1\%; NN: 20.3± 14.9\%, \textit{p}<.05). Fig.\ref{fig:SDNN_results}(b) shows the percent changes in SDNN between S3 and S2. A significant increase in SDNN was observed across the four groups [\textit{F}(3, 36) = 4.416, \textit{p}<.01]. Specifically, in PB group, SDNN increased significantly larger than PN (\textit{p}=.007) and NN (\textit{p}=.039) groups. In addition, there was a statistically significant interaction between the effects of biofeedback and personalization on enhancing SDNN during relaxation session, \textit{F}(1, 48) = 4.153, \textit{p}=.047. The main effect of Biofeedback is significant, \textit{F}(1, 48) = 7.90, \textit{p}=.007. 


\subsubsection{Respiration rate, RES}
\begin{figure}[h]
  \centering
  \includegraphics[width=\linewidth]{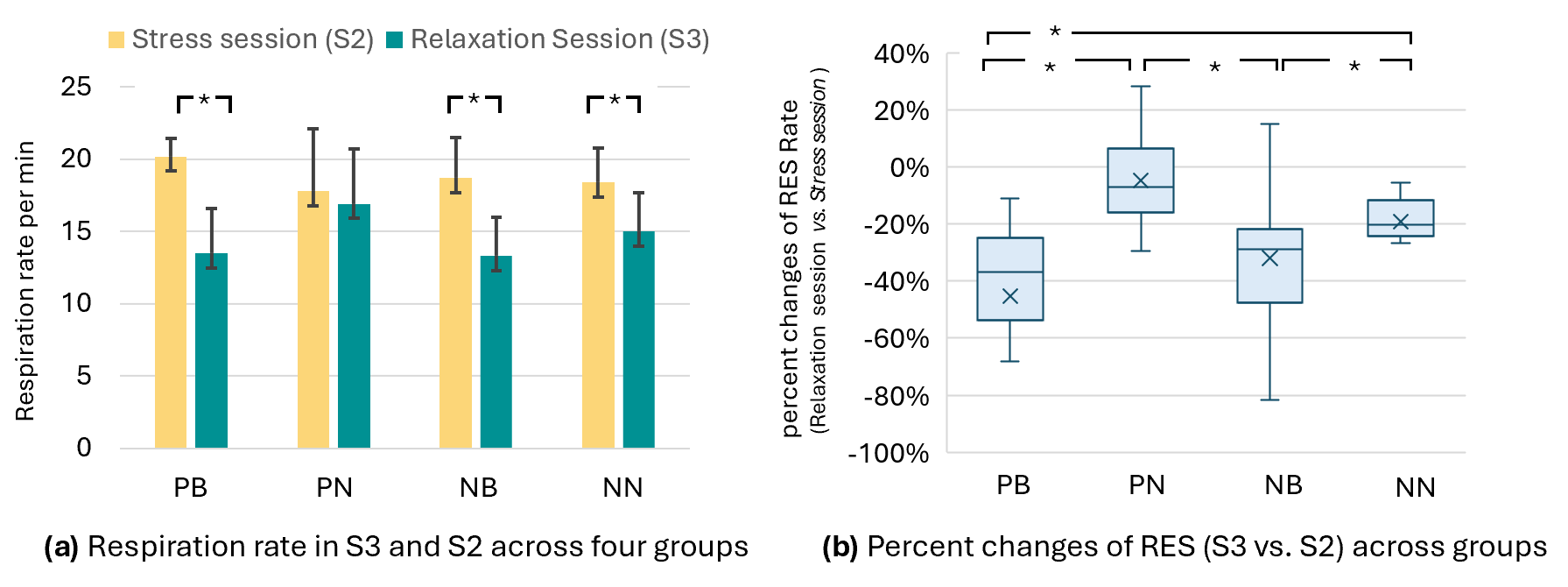}
  \caption{(a). Respiration Rate changes (S3\textit{vs.} S2) across four groups (b). Boxplots of percent changes in RES between S3 and S2}
  \Description[Respiration rate during stress and relaxation session across conditions.]{Two plots compare four groups (PB, PN, NB, NN). The left panel shows average respiration rate (RES) in cycles per minute for the stress session (S2) and the relaxation session (S3), with lower respiration rates in S3 than S2 for all groups. In PB, NB and NN groups, the RES significantly decreased during S3 compared to S2. The right panel shows the percent change in respiration rate from S2 to S3 for each group, with larger decreases in the two biofeedback groups than in the two non-biofeedback groups. There was a statistically significant interaction between biofeedback and personalization.}
  \label{fig:RES_results}
\end{figure}

Fig.\ref{fig:RES_results} (a) shows the average respiration rate (cycle per min) during the relaxation session (S3) and the stress session (S2). Comparing S3 to S2, RES decreased in all groups(PB: -45.2± 35.5\%; PN: -4.6± 15.9\%; NB: -31.9± 22.9\%; NN: -19±17.2\%). And the RES decrease was statistically significant in PB (\textit{p}<.001), NB (\textit{p}<.001), and NN (\textit{p}=.003) groups. Across the four groups, a significant difference in the RES decrease was observed $\chi^2(2) = 20.81, \; p < .001$. The RES decrease was significantly greater in the biofeedback groups (PB and NB) than in the non-biofeedback groups (PN and NN) (\textit{p} <.005). Additionally, there was a statistically significant interaction between biofeedback and personalization, indicating that their combined effect led to a greater reduction in participants' RES\textit{F}(1, 48) = 4.126, \textit{p}=.045.


\subsubsection{Relaxation Rating Scale (RRS)}
\begin{figure}[h]
  \centering
  \includegraphics[width=\linewidth]{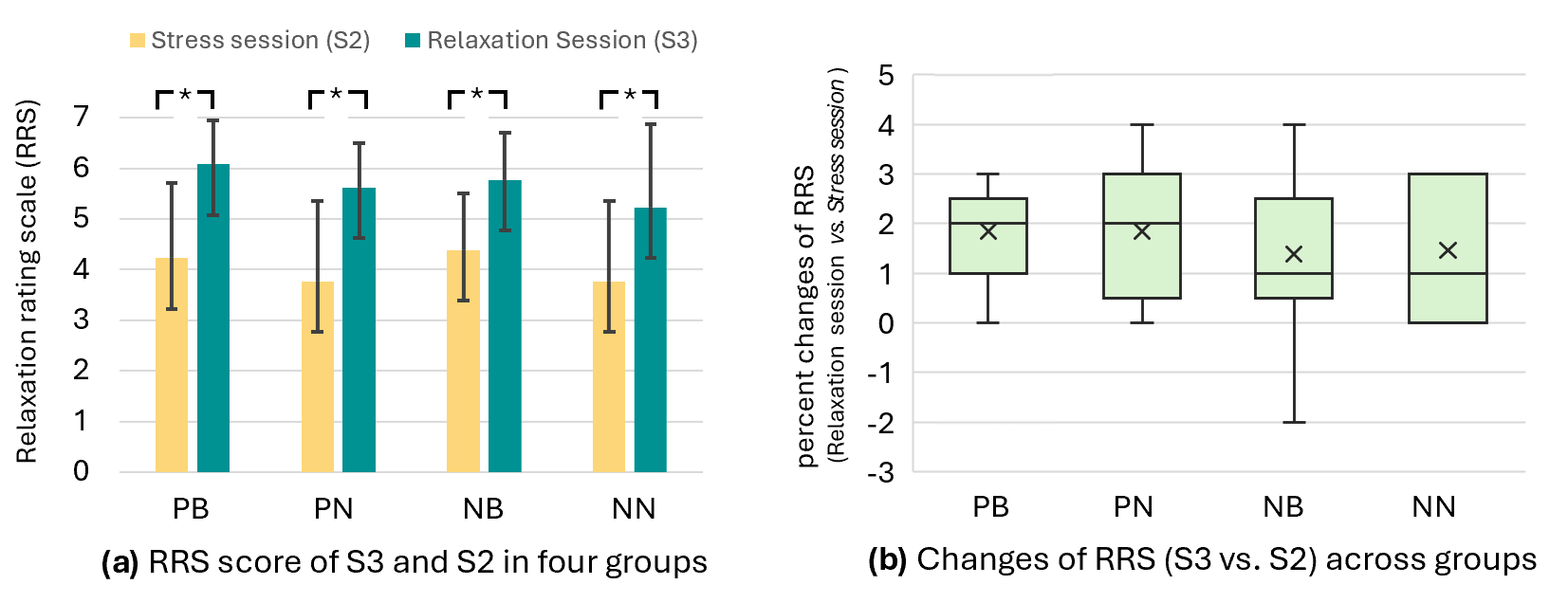}
  \caption{(a). Relaxation rating scale (S3\textit{vs.} S2)  across four groups (b). Boxplots of percent changes in RRS between S3 and S2}
  \Description[Self-reported relaxation ratings during stress and relaxation across conditions.]{Two plots compare four groups (PB, PN, NB, NN). The left panel shows average Relaxation Rating Scale scores for the stress session (S2) and the relaxation session (S3), with higher ratings in S3 than S2 for all groups. In all groups, the relaxation rating scale increased significantly during S3 compared to S2. The right panel shows the change in Relaxation Rating Scale scores from S2 to S3 for each group, indicating similar increases across groups with no clear separation between conditions.}
  \label{fig:RRS}
\end{figure}

As shown in Fig.\ref{fig:RRS} (a), for the participants in all groups, the relaxation rating scale increased significantly during the relaxation session (S3) compared to the stress session (S2)(PB: 1.85± 1.86, \textit{p}=.01; PN: 1.85± 1.46, \textit{p}=.005; NB: 1.38± 1.56, \textit{p}=.017; NN: 1.46± 1.27, \textit{p}=.007). Although the average increase of RRS scores was greater in the PB and PN groups, these differences were not statistically significant compared to NB and NN groups.


\subsubsection{State-Trait Anxiety Inventory, STAI-S}
\begin{figure}[h]
  \centering
  \includegraphics[width=\linewidth]{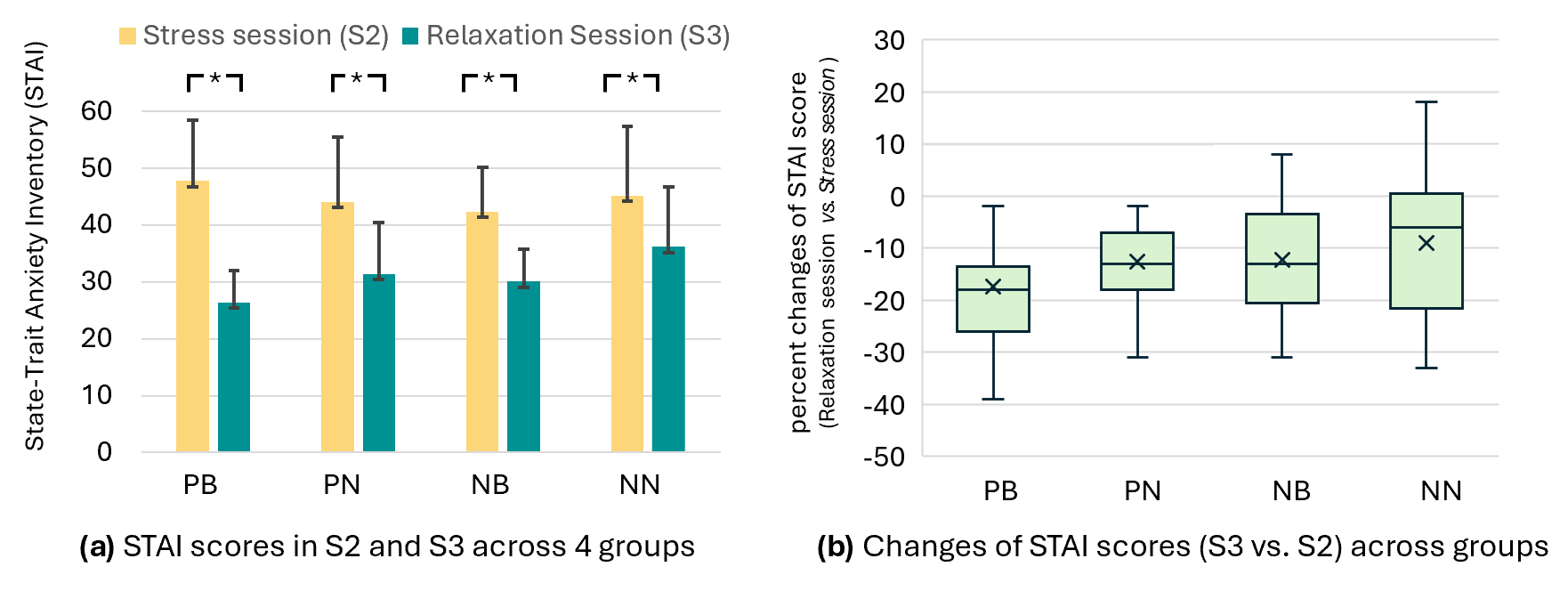}
  \caption{(a). STAI scores (S3\textit{vs.} S2) across four groups (b). Boxplots of percent changes in STAI between S3 and S2}
  \Description[State anxiety scores during stress and relaxation across conditions.]{Two plots compare four groups (PB, PN, NB, NN). The left panel shows average State-Trait Anxiety Inventory–State scores for the stress session (S2) and the relaxation session (S3), with lower scores in S3 than S2 for all groups. For all groups, the STAI scores decreased significantly during S3 compared to S2. The right panel shows the change in State-Trait Anxiety Inventory–State scores from S2 to S3 for each group, indicating decreases in all groups and the largest reduction in the PB group, without a clear separation among conditions.}
  \label{fig:STAI_results}
\end{figure}

As shown in Fig.\ref{fig:STAI_results}, for all groups, the STAI scores decreased significantly during the relaxation session compared to the stress session (PB: -17.38± 11.93, \textit{p}<.01; PN: -12.62± 10.74, \textit{p}=.01; NB: -12.23± 10.78, \textit{p}=.01; NN: -9±14.14, \textit{p}=.04). Although the STAI score shows the largest decreases in the PB group (-17.38± 11.93), there was no statistically significant difference between the four groups (\textit{F}(3,48) =1.082,\textit{p}=.037). 

\subsubsection{Igroup Presence Questionnaire (IPQ) Scale}
\begin{figure}[h]
  \centering
  \includegraphics[width=\linewidth]{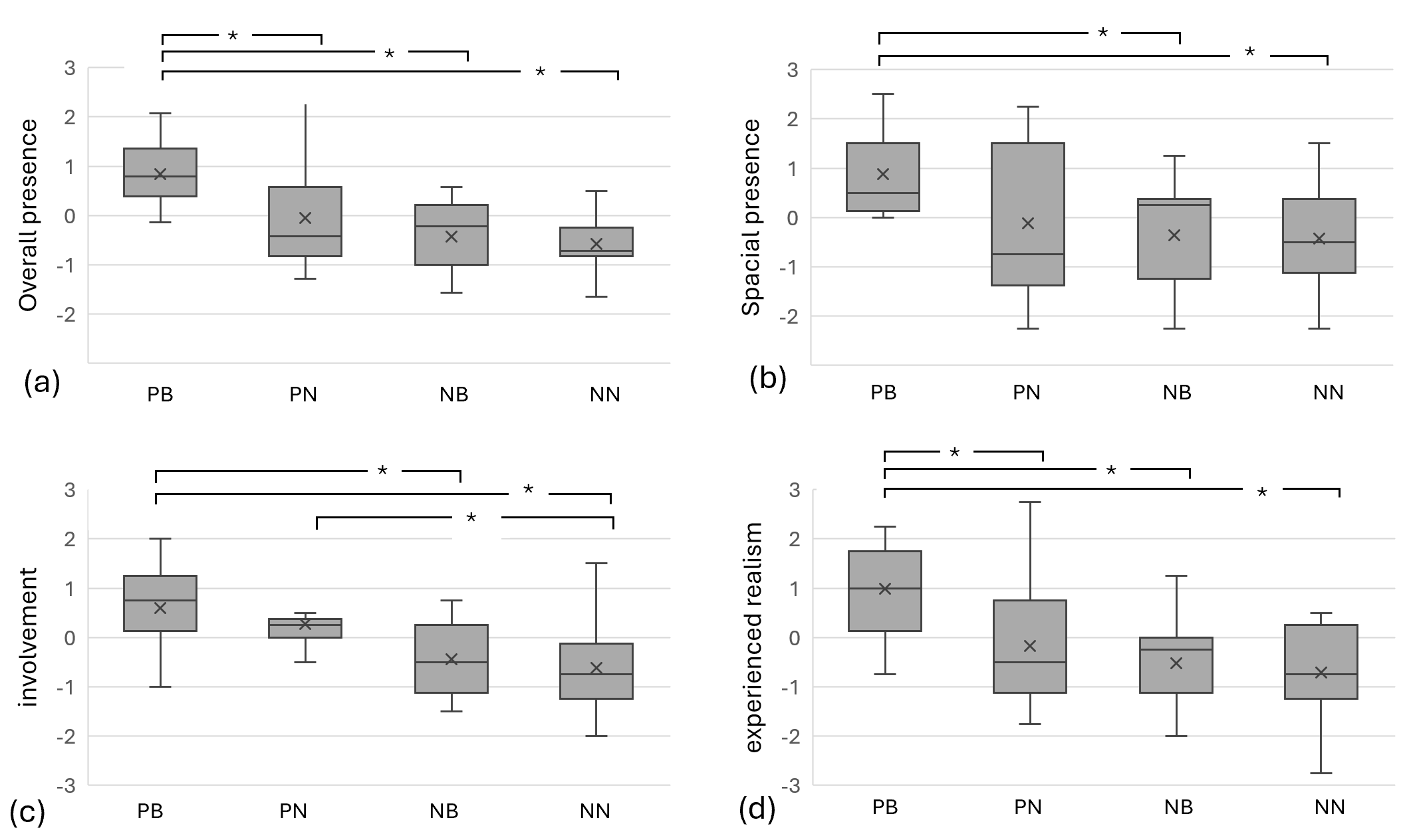}
  \caption{The results of igroup presence questionnaire (IPQ)}
  \label{fig:IPO}
  \Description[Presence questionnaire scores across conditions.]{Four bar charts compare igroup Presence Questionnaire ratings for four groups (PB, PN, NB, NN). The top-left panel shows overall sense of presence, where the personalization-with-biofeedback group has the highest score and is higher than the other three groups. The top-right panel shows spatial presence, with higher scores for the personalization groups than the non-personalization groups. The bottom-left panel shows involvement, where both personalization groups score higher than the no-personalization groups. The bottom-right panel shows experienced realism, with the personalization-with-biofeedback group highest and higher than the other three groups.}
\end{figure}

Fig.\ref{fig:IPO} shows the results of the igroup presence questionnaire (IPQ) between the four groups. As shown in Fig.\ref{fig:IPO} (a), the\textit{overall sense of presence} of PB group (0.84 ± 0.64) is significantly higher than PN (-0.05 ± 0.99, \textit{p}=.019), NB (-0.43 ± 0.69, \textit{p}<.001), and NN (-0.58 ± 0.59, \textit{p}<.001) groups. A two-way ANOVA analysis shows a significant main effect of personalization on \textit{sense of presence}, \textit{F}(1, 48) = 18.58, \textit{p} =<.001. The main effect of Biofeedback on \textit{sense of presence} was also significant, \textit{F}(1, 48) = 6.311, \textit{p} = .015. However, no significant interaction was found between biofeedback and personalization, \textit{F}(1, 48) = 3.23. \textit{p} =.078. This result indicates that both biofeedback and personalization via \general{art-therapy-inspired activity} contribute independently to enhance the sense of presence during VR relaxation sessions, but they do not amplify or weaken each other's effects.

As shown in Fig.\ref{fig:IPO} (b), the ratings on \textit{spatial presence} in the PB group (0.88 ± 0.88) is significantly higher than the NB (-0.36 ± 1.04, \textit{p}=.043) and NN (-0.42 ± 1.01, \textit{p}<.031). A significant main effect of personalization on \textit{spatial presence} was observed, \textit{F}(1, 48) = 5.78, \textit{p} =.02. This indicates that the participants (in the groups of PB and PN) who received a personalized 360 environment and relaxation guidance would perceive a higher sense of spatial presence (M = 0.38, SD = 0.27) than those who did not (in the groups of NB and NN) (M = -0.39, SD = 1.0). As shown in Fig.\ref{fig:IPO} (c), for the sub-scale \textit{involvement}, PB group (0.59 ± 0.86) scored significantly higher than NB (-0.44 ± 0.74, \textit{p} = .013) and NN group (-0.61 ± 0.95, \textit{p} = .003). PN group (0.27 ± 0.74) scored significantly higher than the NN group (\textit{p} =.044). A significant main effect of personalization on sense of involvement was observed, \textit{F}(1, 48) = 17.4, \textit{p} <.001, indicating that the participants (in PB and PN groups) performing relaxation with personalized environment and guidance perceive a higher sense of involvement (M = 0.43, SD = 0.8) than those who did not (NB or NN group) (M = -0.52, SD = 0.84).


As shown in Fig.\ref{fig:IPO} (d), the scores of \textit{experienced realism} in group PB (0.98 ± 0.96) is significantly higher than that of the groups of PN (-0.17 ± 1.23, \textit{p}=.03) , NB (-0.52 ± 0.91, \textit{p}<.003), NN(-0.71 ± 0.97 , \textit{p}<.001).  A significant main effect of personalization \textit{F}(1, 48) = 12.8, \textit{p} =<.001. and a significant main effect of Biofeedback \textit{F}(1, 48) = 5.9, \textit{p} =.02. were observed. There was a trend toward an interaction between biofeedback and personalization, suggesting that they may amplify each other’s effects in enhancing the \textit{experience of realism}; however, this interaction did not reach statistical significance, \textit{F}(1, 48) = 2.85. \textit{p}=.09.


 

\subsubsection{Flow State Scale (FSS) and Perceived relevance}
\begin{figure}[h]
  \centering
  \includegraphics[width=\linewidth]{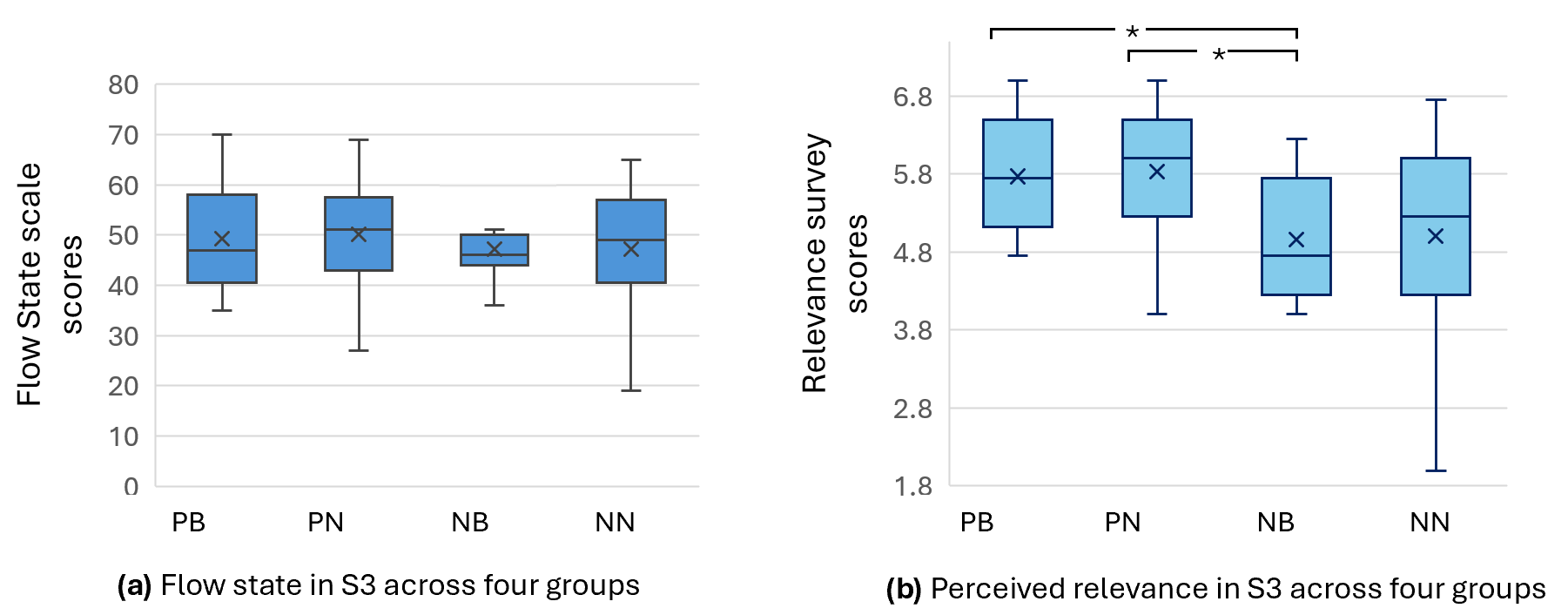}
  \caption{The results of (a) Flow State Scale and (b) Perceived relevance}
  \Description[Flow state and perceived relevance ratings during the relaxation session across conditions.]{Two plots compare four groups (PB, PN, NB, NN) in the relaxation session (S3). The left panel shows Flow State Scale scores, which are similarly high across all groups with no visible separation. The right panel shows perceived relevance ratings, where the two personalization groups report higher relevance than the two non-personalization groups, with the no-personalization-with-biofeedback group lowest. The perceived relevance score was significantly higher in PB and PN groups than the NB group.}
  \label{fig:FSS_Relevance}
\end{figure}

As shown in Fig.\ref{fig:FSS_Relevance} (a), all groups reported a high flow state during the relaxation session (PN: 49.31: ± 10.9; PN: 50.00± 12.03; NB: 47.15± 7.15; NN: 47.15± 13.7). There is no significant difference between the groups. Fig.\ref{fig:FSS_Relevance} (b) shows the participants' \textit{perceived relevance} between four groups. A statistically significant between-group difference was observed (\textit{F}(3, 48) = 2.98, \textit{p} =.04). The perceived relevance  score was significantly higher in PB(5.77 ± 0.75, \textit{p}=.014) and PN (5.83 ± 0.89, \textit{p}= .016) groups than the NB group (4.96 ± 0.8).  In addition, there was a significant main effect of personalization (\textit{F}(1, 48) = 8.9  \textit{p}= .004). Participants (in PN and PB) receiving personalized relaxation environment and guidance reported a higher perceived relevance about the relaxation session (M = 5.79, SD = 0.80) than those (in NB and NN) who did not (M = 4.98, SD = 1.1). 




\subsection{Qualitative Findings}
In this section, we illustrate our findings with interview excerpts and recordings of the \general{art-therapy-inspired activity} sessions. We retraced 26 participants’ (PB and PN group) sessions and reviewed the sketches and environment maps they created, reflecting on recurring patterns and trends. Based on the qualitative results, six themes were derived from the interviews: \textit{Forms of Safe Place Imagery for Security and Relaxation}, \textit{Inspirations Behind Safe Place Creation}, \textit{Familiarity and Memory in Personalized Safe Places}, \textit{Enhancing Safe Places through Personalized Audio Guidance}, \textit{Connecting Body and Environment through Biofeedback}, and \textit{Additional User Preferences in Safe Place Creation}.

\subsubsection{Safe Place Imagery for Security and Relaxation}
\begin{figure*}[h]
  \centering
  \includegraphics[width=\linewidth]{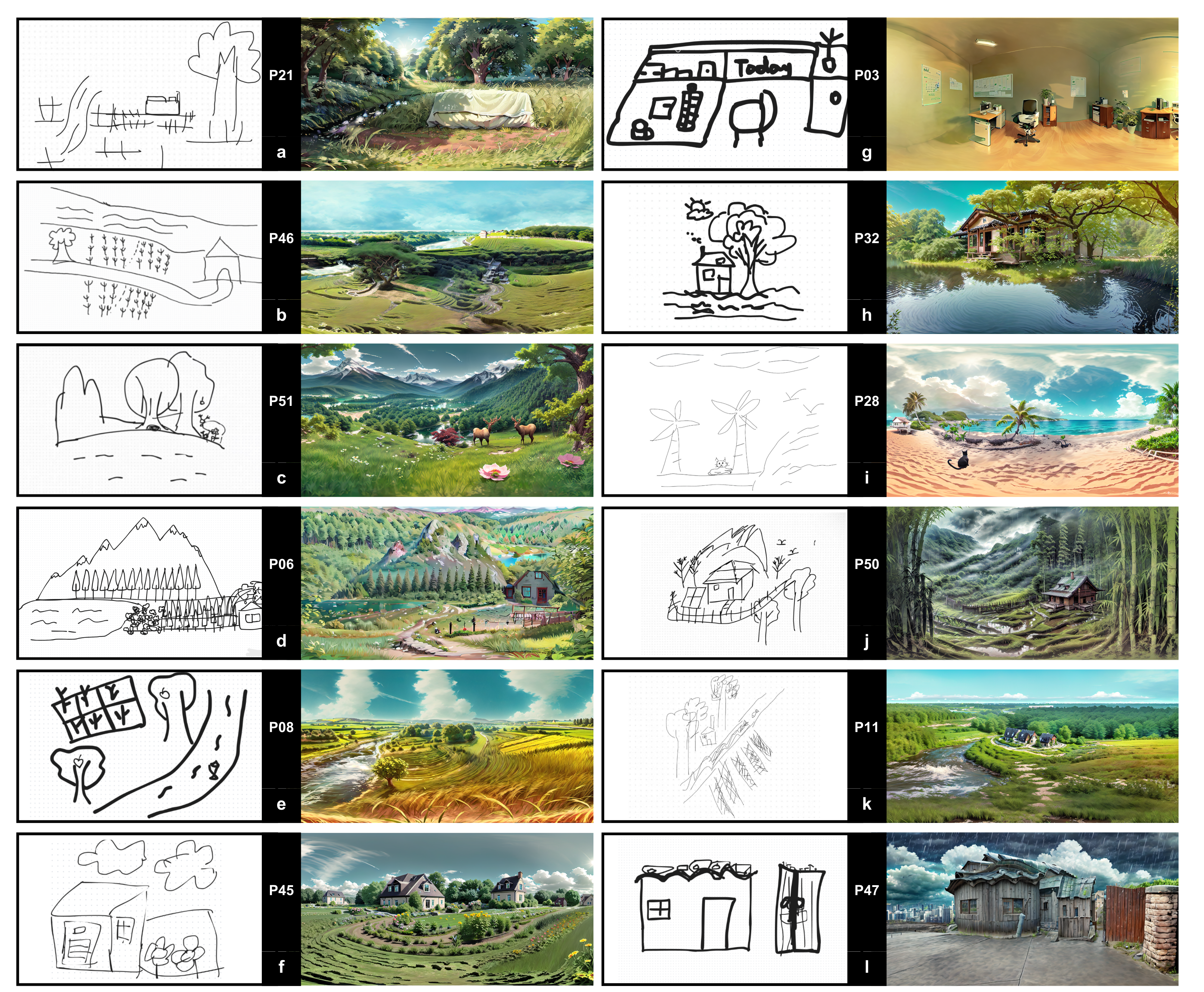}
  \caption{The example of environment maps and the corresponding sketches depicted by the user during the personalization phase.}
  \Description[A grid illustration showing examples of environment maps and the corresponding sketches depicted by the user.]{Figure 11(a) shows the example depicted by P21: a bed in the forest. Figure 11(b) shows the example depicted by P46: a field near a river with a road to a house. Figure 11(c) shows the example depicted by P51: a lake surrounded by mountains with deers hanging around. Figure 11(d) shows the example depicted by P06: a house down to the mountains, near a lake. Figure 11(e) shows the example depicted by P08: golden fields with apple trees near a river. Figure 11(f) shows the example depicted by P45: houses with gardens. Figure 11(g) shows the example depicted by P03: office area with their daily objects. Figure 11(h) shows the example depicted by P32: a house and a tree beside the water. Figure 11(i) shows the example depicted by P28: a black cat on the beach with blue sky and the sea. Figure 11(j) shows the example depicted by P50: a house with wooden fence in bamboo forests. Figure 11(k) shows the example depicted by P11: a house surrounded by fields and forests near a river. Figure 11(l) shows the example depicted by P47: a building with metal roof and iron gate. All the generated environment map are selected and modified by the corresponding participant and reflects key elements in their safe place.}
  \label{fig:exampleoutput}
\end{figure*}

Participants created a wide variety of safe places, with few resembling one another, highlighting the highly individualized nature of these imagined environments. Despite this diversity, clear patterns emerged across the data. Most safe places were set in natural and outdoor environments, such as forests (P02, P05, P06, P11, P21, P39, P50) and seaside (P04, P23, P25, P28, P48), while only a few depicted indoor settings. For example, one participant (P21) imagined a cozy bed in a forest near a brook, whereas another (P03) envisioned the interior of an office room (see Fig.\ref{fig:exampleoutput} (a) and (g)).

Plants appeared in most participants’ safe places. Furthermore, some participants depicted specific types of plants, for example, a huge banyan tree (P46) or tall trees covered with fragrant osmanthus blossoms, with fallen petals carpeting the ground (P32; see Fig.\ref{fig:exampleoutput} (b) and (h)), which highlights both the demand for and the potential of personalization.
  
Similarly, houses were a common element (e.g., a house enveloped by enormous trees, P06; a Chinese-style building in a bamboo forest under the mountain, P50; see Fig.\ref{fig:exampleoutput} (d) and (j)), possibly suggesting a recurring motif of privacy, shelter, and safety. Common elements also included rivers and lakes, as well as mountains.

Several participants depicted animals. Five of them mentioned wild animals in the environment (P51, P28, P05, P50, P48), while another five included pets they kept (P51, P48, P28, P14, P44). For instance, P51 asked the researchers to add reindeers into her safe place during the modification, as the initial version did not include the reindeers she had depicted in her sketch. She explained: "The animals felt like my guardian spirits — companions or even mounts and pets — whose presence gave me a sense of being protected" (see Fig.\ref{fig:exampleoutput} (c)). By contrast, P28 said there was no need to modify the black cat in the initial version of her environment map, even though she had depicted an orange cat. She explained: "We had two cats — one black and one orange. At the time, I felt the orange one fit better with the sunlight, so I chose the orange, though the black one would be fine too" (see Fig.\ref{fig:exampleoutput} (i)). These examples illustrate how participants actively personalized their safe places, assigning symbolic meaning to animals and plants, which were not merely decorative but functioned symbolically as guardians, companions, or representations of personal memory. Their inclusion reinforced the participants’ sense of protection, familiarity, and emotional connection to the environment.

Five participants mentioned fields and crops (e.g., a rice field with its stalks crowned with golden ears of grain, P08; a field where I can plant some crops, P11; see Fig.\ref{fig:exampleoutput} (e) and (k)), which may reflect their sentimental attachment to farming and rural life, and show how they incorporate personally meaningful elements into their safe places.

\subsubsection{Inspirations Behind Safe Place Creation}
Most participants incorporated their memories into their safe place visualizations, while only a few constructed their safe places purely from imagination or inspired by films. For example, P51 imagined a safe place named \textit{The Spirit World} (see Fig.\ref{fig:exampleoutput} (c)), and she elaborated:” \textit{Sometimes I imagine myself as a witch. I feel a strong sense of security when I can use magic to control the world. The image of the witch is probably influenced by films and TV shows. One character I really like is Wanda, the Scarlet Witch from the Marvel universe}”.

Among the safe places inspired by memories from participants’ own lives, some were rooted in childhood memories. These memories could be related to memorable places (e.g., grandmother’s house which has been changed and is no longer in its original state; P35, safe place named \textit{A Pastoral Landscape Painting}), personal experiences about themselves (e.g., in middle school days, classes were canceled on days of rainstorms; P42, safe place named \textit{A Forest Veiled in Mist}), or shared experiences with family and friends (e.g., go on a seaside vacation with family; P28, safe place named \textit{Sunny Beach}; see Fig.\ref{fig:exampleoutput} (i)).

Beyond childhood, experiences from adulthood also served as an important source of inspiration for some participants. These also included personal experiences, such as the impressive scenery seen during a trip to Norway (P14, safe place named \textit{Merge My Light With the Dust of the World}), as well as shared experiences with family and friends, for example, a trip to a farm with a friend (P05, safe place named \textit{A Sunny and Warm Place}).

These patterns imply that autobiographical experiences are a primary source of inspiration for imagining environments that evoke security and relaxation. Though fantasy-based safe places can be created using our system and are effective for some participants, memory-based personalization may offer deeper relevance and meanings.

\subsubsection{Familiarity, Pleasant Surprise, and Memory Retrieval in Personalized Safe Places}
In the \general{art-therapy-inspired activity} session or the interview after the relaxation session, twelve participants explicitly reported that their safe place aligned with their imagination. In analyzing participants’ responses to the open-ended interview questions about their relaxation experiences, four participants highlighted the sense of familiarity evoked by their personalized safe places (P28, P46, P32, P49). For example, P28 thought the personalized audio guidance “\textit{sounded quite familiar and I felt a sense of relaxation}”. Similarly, P49 identified that the audio guidance “\textit{was designed for me and thus familiar and amiable}”.

In addition to familiarity, participants also reported that the personalized safe places gave them pleasant surprise (P51, P42, P44, P45, P47, P4). The pleasant surprise mainly arose from the differences or supplements that AI introduced in the personalized content generation. For example, P45 said “\textit{It matched what I had in mind, and it added a lot more — like a seaside, many trees, and lots of houses, so there are neighbors and I wouldn’t feel lonely}” (see Fig.\ref{fig:exampleoutput} (f)). P51 thought there is a pleasant surprise in both the environment map and guidance audio: “\textit{I felt that the picture was actually not quite the same as what I had imagined before, but I still liked it. (...) Hearing my earlier imagination and descriptions being presented in such a ceremonial way gave me a sense of surprise}” (see Fig.\ref{fig:exampleoutput} (c)).

Another interesting finding is that our system provided an experience of “retrieval”: the generative AI’s capabilities can recreate scenes or objects that no longer exist, allowing users to revisit and reflect on their memories during the co-creation process. For example, P47 found the safe place unexpectedly amazing: ” \textit{I actually thought it was pretty amazing, because these are memories from when I was very young, before I even started school, so they’re actually quite vague. I had been looking forward to seeing that place again, but it seems it’s already been flattened and replaced with buildings, so I can never see it in real life. In this AI creation process, it felt a bit like having the things I deeply wanted to see presented to me again. At first, my expectations of the AI weren’t that high — I didn’t think it could generate what I wanted, especially since the images were so rough. But in the end, what it generated was actually even better than I had expected}” (see Fig.\ref{fig:exampleoutput} (l)).

\subsubsection{Imagining with Sensation, Role-Playing and Narration through Personalized Audio Guidance}
Participants reported that they feel relaxed when imagining with the personalized audio guidance: ”\textit{When I closed my eyes, I felt a pleasant outward expansion. It was so comfortable that closing my eyes made it easy to imagine and follow the guidance to explore further}” (P48), “\textit{It guided me to think about things that made me feel good}” (P05), “\textit{I started to feel very relaxed after hearing the guidance. I thought of the kind of relaxation I feel when fishing, touching that tree, or walking along a path — the feeling of calm slowly emerged}” (P46), among others.

We looked at the script of these personalized guidance audio, and found it guided participants to move or to feel the senses beyond the purely visual, incorporating auditory, tactile, and olfactory elements to create a more embodied experience. For example, the guidance invited participants to “\textit{reach out to touch the flowers and plants, their soft textures seem to respond to your breath, and each leaf carries a gentle, fuzzy tenderness}” (P48) or “\textit{place the hand on the rough bark of the banyan tree, you felt a weighty, real texture that seemed to let me perceive the years etched into it and the warmth of the soil beneath}” (P46). Furthermore, the guidance introduced motion to otherwise static visual elements, and highlighted naturally dynamic phenomena, including the gentle rustle of wind (P03), the soft dispersal of flower fragrance (P10), and the sky dyed red by the sunset (P05), enhancing the sense of a living environment.

Finally, role-playing and narrative elements were included to promote personal identification and memory recall: “\textit{As you once mentioned, sometimes you imagine yourself as a witch, gently closing your eyes in the magic circle, practicing magic, or quietly spending time with your spiritual animal companions — the spirit world itself is your sanctuary}” (P51), and “\textit{In childhood, there were moments like this: basking in the sun, feeling the wind, with a dog lying at your feet; that familiar warmth resurfaces in your ‘Carefree Island’}” (P48).

Together, these audio-guided enhancements enriched the safe place experience by supplementing and extending the sensory and dynamic qualities that static images alone could not convey.

\subsubsection{Breath–Environment Synchrony and Body–Environment Connection through Biofeedback}
We analyzed all participants’ (52 in total) responses to the question “Do you notice the particle effects and how do you feel about it?” As a result, most participants in AB and NB group who experienced relaxation training with biofeedback liked the particle effects, while some participants in AN and NN did not notice the effects or they did not have specific impressions.

Some participants from the AB and NB groups felt that the particle effects made their safe places more dynamic and, at times, blended naturally into the environment. For example, some described the moving particles as resembling the flow of a river (P11), the waves of the sea (P14), or grass swaying in the wind (P35).

Several participants (P48, P25, P22, P35, P9, P38, P43) also felt a resonance and connection between their bodies and the environments: “\textit{This breathing sensation gives me a slight sense of presence, as if it moves in rhythm with my breath. From the perspective of presence, I feel it creates a complete visual experience along with a certain tactile presence}” (P35), “\textit{I could follow her rhythm of contraction, resonating in sync, like a kind of spiritual energy in meditation}” (P25), “\textit{I felt that the scene was connected to me, as if I could somehow take control of it}” (P43).

\subsubsection{Additional User Preferences in Safe Place Creation}
In addition to the broader themes described above, participants also expressed some more specific preferences when depicting their safe places.

For auditory elements, many participants highlighted birdsong and the trickle of water. For olfactory elements, most participants prefer fresh smells, no matter it came from plants (e.g., the scent of flowers mixed with the smell of fresh grass; P14), soil (e.g., the earthy freshness that comes when rain soaks into the ground; P46) or water (e.g., the fresh, clean scent of mountain spring water; P08). This preference may reflect that such fresh smells help participants relax, although it could also be influenced by the fact that the user study was conducted in the hot summer season. Similarly, nine participants mentioned soft and furry textures when asked to imagine haptic sensation of the elements in their safe place: ”\textit{when I touched the reindeer, its fur was fluffy and cozy, giving me a sense of security}” (P51), “\textit{the grass and my cat felt soft and furry” (P44), “the sand is very soft}” (P23). These soft and furry textures appeared to provide participants with a sense of security and relaxation.

Common patterns also included mild sunshine that was “not too hot, just warm enough to be comfortable” (P08, P02, P42, P25, P23). While these preferences varied from person to person, they reflected patterned tendencies in how participants infused their safe places with personal touches, often drawing on gentle, natural, and soft cues to enhance the sense of safety and familiarity.

\section{Discussion}

Previous work on biofeedback has demonstrated its benefits for stress management and relaxation training \cite{3,5,kennedy2019biofeedback}. While personalization has been recognized as a key factor in enhancing the effectiveness of biofeedback systems for relaxation training, 
existing approaches have rarely examined the integration of \general{an art-therapy-inspired activity} as a means of personalizing 360° virtual environments and audio guidance within relaxation training. Even fewer studies have examined how users' imagination of a safe and relaxing place, captured through verbal and non-verbal expressions, along with personal experiences, familiar symbols, and memorable emotions and sensations can be leveraged as input to deeply personalize the relaxation training experience. In this study, we explore leveraging the user’s personal life experiences derived from expressive art therapy to develop a personalized calming 360 environment and relaxation guidance as the user interface of a VR biofeedback system for relaxation training. To answer RQ1, we compared physiological and subjective relaxation outcomes across four conditions (i.e., biofeedback with personalization, biofeedback only, personalization only, and neither) to examine whether art-therapy-derived personalization can complement biofeedback in jointly enhancing relaxation training. To answer RQ2, we derived design implications and future directions by thematically analyzing interviews and \general{art-therapy-inspired activity} sessions with participants who experienced relaxation training with art-therapy-derived personalization.




\subsection{Personalization enabled by \general{art-therapy-inspired activity} and biofeedback jointly support relaxation training}
\textbf{Significant interaction between biofeedback and personalization.} 
Our results confirmed the feasibility of combining biofeedback with personalization enabled by \general{art-therapy-inspired activity}, demonstrating synergistic effects in improving HRV and reducing respiration rate. Specifically, we observed a significantly higher increase of HRV-SDNN in the PB group, which suggests that combining biofeedback with personalization produced positive effects for enhancing autonomic regulation \cite{von2007heart}. While many prior studies \cite{lehrer2007biofeedback,lehrer2020heart} have shown biofeedback alone can enhance HRV by training individuals to regulate stress responses, our findings about the significant interaction between biofeedback and personalization suggest personalization may further amplify the efficacy of biofeedback. In response to \textbf{RQ1}, these results confirm that personalization and biofeedback can mutually reinforce each other in relaxation training. The qualitative results further explained and supported this finding. Personalized ‘safe places’ environment and audio relaxation guidance enhanced user engagement and relaxation, while biofeedback amplified their effectiveness by providing real-time guidance during breathing exercises.


\textbf{Fundamental role of biofeedback as a self-regulation tool.} This study further confirms the evidence of previous studies \cite{whited2014effectiveness, yu2018biofeedback, yu2018delight} on biofeedback for relaxation training. In both biofeedback groups (PB and NB), significant changes were observed between the stress and relaxation sessions, including increased HRV, decreased breathing rate, reduced STAI scores, and increased RRS scores.  In particular, participants in the biofeedback groups (PB and NB) achieved significantly slower respiration rates compared to those in the non-biofeedback groups (PN and NN). This suggests that biofeedback played a direct role in supporting breathing exercise, enabling participants to become more aware of their breathing patterns and self-regulate them in real time. The slowed respiration rate is a well-established indicator of relaxation and parasympathetic activation, highlighting the effectiveness of biofeedback in promoting calmness and stress reduction. Importantly, this effect was observed regardless of whether personalization was included, underscoring the fundamental role of biofeedback as a physiological training tool. 

\textbf{Personalization via \general{art-therapy-inspired activity} strengthening user involvement and meaningfulness.} Our findings demonstrate that personalization derived from \general{art-therapy-inspired activity} significantly enhanced both participants' \textit{sense of involvement} and their \textit{perceived relevance} of the relaxation sessions. Participants who engaged in the personalization phase through \general{an art-therapy-inspired activity} (PB and PN) reported substantially higher levels of involvement compared to those who did not (NB and NN), suggesting that the outcome features of personalized 360 'safe place' environment and life-experience infused audio guidance could foster a stronger connection to relaxation practice. Also, the higher \textit{perceived relevance} ratings among PB and PN groups suggest that personalization made relaxation sessions feel more contextually and personally meaningful. This aligns with prior research \cite{52, 33} on the personalization of relaxation practices and extends it by demonstrating a proof-of-concept for using \general{art-therapy-inspired activity} to incorporate personal experiences and emotional preferences into a highly personalised relaxation training approach.

\subsection{Design Implications}

\subsubsection{\textbf{Design Relaxation Environments with Personally Meaningful Symbols, Memory-Evoking Imagery, and Familiar Sensory Cues}}


Our qualitative findings revealed recurring patterns in how participants conceptualized and engaged with their personalized safe places, highlighting the following three practical strategies to design personally meaningful relaxation environments.

\textit{Incorporate symbolic cues to evoke feelings of safety and comfort.} In our user study, although the safe places envisioned by participants were highly individualized, certain elements recurred. The overlap of these elements suggests that users may attach subtle symbolic meanings to them, for example, houses as symbols of shelter, animals as companions, or fields as representations of farming life. Importantly, such symbols do not always require deeply private autobiographical input \cite{440}. As demonstrated in our study, even imagined safe places can be effective as long as they establish a meaningful connection between the user and the environment and evoke symbolic resonance. Research have shown that symbolic and natural elements in spaces can enhance comfort, emotional well-being, and perceived safety \cite{439}. Designers could integrate these motifs subtly into the environment, for instance, through buildings, animals, or plants, so that they evoke the intended symbolic associations without directly replicating the user’s exact memories or imagination.

\textit{Incorporate memory-evoking cues to strengthen emotional connection.} Several participants drew on memory-linked imagery, such as childhood scenes or memorable trips, highlighting the potential of designs to support self-referential memory cues and strengthen the emotional connection between user and environment. Such cues do not need to replicate users’ imagined memories in full detail; instead, approximate or suggestive features can be sufficient to trigger memory and emotion (e.g., familiar everyday objects \cite{441}, or lifetime-familiar cues \cite{442}). In some cases, users may anchor their emotions to past places or memories that can no longer be physically revisited. Embedding such symbolic anchors into relaxation environments may facilitate memory retrieval and deepen emotional engagement \cite{443}.

\textit{Incorporate familiar sensory cues to enhance feelings of presence.} In our study, participants often describe their safe places with sensory details such as warm weather or the trickle of water, indicating that such sensory details enhanced their feelings of comfort and presence. This is consistent with prior research showing that multi-sensory cues intensify presence in virtual environments \cite{444} and that sensory imagery in autobiographical memory retrieval fosters vividness and emotional comfort \cite{445}. Designers might integrate such cues as subtle ambient visual features corresponding to the user’s imagined sensations (e.g., mild warm sunshine, flowing water, or raindrops), evoking a sense of familiarity and further fostering presence.

\subsubsection{\textbf{Integrating Verbal and Non-Verbal Inputs to Derive Complementary and Mutually Transferable Personalization Cues}}

A persistent challenge in personalization for relaxation training lies in how to elicit and translate users’ lived experiences into meaningful design elements \cite{36}. Recent research has explored the use of LLMs for personalization \cite{30}. Our study demonstrates that combining verbal reflection with non-verbal expression can help uncover richer, more personal cues, which are complementary and mutually transferable.

\textit{Combine enriched personal inputs to articulate meaningful elements and experiences.} Through the Safe Place exercise, participants described environments associated with safety and relaxation. At the same time, participants’ sketches provided complementary insights that were not easily articulated verbally (e.g., terrain of a remembered place, arrangement of an office space). Such visual Input became particularly powerful when transformed into immersive environments, illustrating how non-verbal forms of disclosure offer unique, irreplaceable dimensions for personalization. In addition to drawing, art-making activities such as sketching, collages \cite{53, 401}, or making models \cite{36, 401, 402} also enable participants to externalize abstract feelings and associations that may not surface through verbal descriptions. Beyond these active forms of self-disclosure, verbal and non-verbal personalization input could also be derived from passively collected personal data, such as digital journals, photo archives, or even social media posts \cite{403}. These resources often contain highly intimate traces of users’ lived experiences and emotional needs. Because they reflect individual preferences, memories, and sources of comfort, such data could provide another powerful pathway for identifying input that resonate with users’ need for security and relaxation. 

\textit{Translate diverse personalization input across multiple modalities.} In \name, the personalized environment was not generated solely from the user’s sketch; rather, the user’s detailed descriptions could also serve as prompts for generation. Users were further able to provide instructions to modify the generated map, meaning that their verbalization directly influenced the personalization of visual content. For example, the beautiful colors, mild lightings and even atmospheres were often absent in sketches but were described verbally, and many participants reported feeling comfortable with these visuals in their imagined safe places. While we did not inversely use sketches to guide the personalization of audio guidance, future designs could explore how multimodal personalization input interact, particularly when multiple rich sources of input are available. 

\subsubsection{\textbf{Leverage Sensation-Related Audio Guidance to Compensate for Limited Sensory Modalities}}

Immersion and presence remain central challenges in the design of effective relaxation training systems \cite{12}. Most XR relaxation systems rely primarily on audio and visual stimuli, with few incorporating haptic (touch) or kinesthetic (movement) information, largely due to hardware and implementation constraints. Our findings highlight the potential of personalized audio guidance to address these limitations by enriching the limited sensory feedback typically available in virtual environments. For instance, audio can guide users to imagine approaching a virtual river depicted in their artwork, feeling the gentle flow of water or smelling the scent of rain, thereby evoking related memories and strengthening their personal connection with the environment. Users may close their eyes during relaxation exercises, as guided imagination plays a critical role in enriching the environment and compensating for the absence of other sensations. Furthermore, diverse forms of audio guidance could be explored to expand users’ sensory imagination. Advances in interactive speech synthesis and adaptive sound generation create opportunities for dynamic, responsive forms of guidance that adjust to users’ states in real time. 

\subsubsection{\textbf{Extract Personally Meaningful Elements from Autobiographical Memory to Derive In-Depth Personalization}}

Our study demonstrates that autobiographical memories, elicited through \general{art-therapy-inspired activity}, can serve as a rich source of personally meaningful elements for personalization. Accessing users’ memories often reveals deeper and more nuanced information than direct questionnaires or interviews, which may fail to capture implicit or emotionally significant details \cite{446, 447}. By guiding participants to externalize memories and associations through sketches and therapeutic dialogues, we were able to translate their autobiographical memories into features that meaningfully enhanced personal relevance. This approach represents a largely unexplored pathway for personalization, distinct from traditional personalization cues such as user preferences or contexts. Large language models (LLMs) offer new opportunities for achieving in-depth personalization, due to their powerful capacity to extract, interpret, and generate information from diverse forms of user input \cite{448}. This capability enables systems to translate users’ autobiographical memories into personally meaningful elements, which can serve as in-depth personalization cues. Besides, such memory-driven personalization can be leveraged in broader potential applications beyond relaxation training. For instance, it could be used to support the current personalization techniques in palliative care \cite{449}, educational simulations \cite{450}, or interactive storytelling experiences \cite{451} to increase user engagement and emotional comfort. 

\subsubsection{\general{\textbf{Build Safeguards to Protect Memory Integrity, Emotional Safety, and Autobiographical Privacy}}}
\general{The use of generative AI to translate autobiographical content into VR imagery raises some potential ethical risks. First, AI models can distort or aesthetically ‘beautify’ personal memories, leading users to blend generated scenes with their original recollections. This is particularly sensitive for childhood, family, or emotionally charged places, where visual details carry personal significance. To reduce such risks, we explicitly and implicitly emphasize to the users that their ‘safe places’ are spaces constructed based on their imagination rather than reproductions to their memories. We also use a stylized image generation model instead of realistic image generation models, creating a clear perceptual boundary between the user’s lived memories and the AI-generated imagery.}

\general{Second, externalizing private memories into visually rich VR scenes can intensify emotional reactions, leading to unexpected associations or mixed emotions. While we employed therapist supervision to mitigate this risk, unsupervised deployments would require safeguards such as constraints in prompts and options to revise or reject AI-generated interpretations.}

\general{Third, personalization pipelines using therapeutic-style dialogue may inadvertently preserve sensitive autobiographical cues, even without explicit identifiers. Emotional tone, relational patterns (e.g., “my father used to…”), culturally specific references, or narrative structure can act as implicit identifiers in model prompts. Modern generative models can also retain semantic traces of these cues within intermediate representations, which means that an apparently anonymized prompt may still encode highly personal meaning \cite{Li_Xu_Song_2023}. This makes anonymization insufficient on its own. To minimize this risk, future systems should incorporate design-layer protections such as filtering relational or emotionally loaded context and performing on-device processing to prevent server retention. Such measures are essential if the system is to support safe, self-guided use beyond controlled research settings.}


\subsection{Limitations and Future Work}
Further work will explore integrating multi-modal interfaces and biofeedback into our system. One limitation of our system is the narrow range of feedback and guidance types. The current design relied mainly on visual and auditory cues, without tangible feedback or background sounds such as music or nature ambience. While this simplification allowed us to focus on core interactions, prior research \cite{47} shows that subtle auditory layers can strongly influence mood, immersion, and relaxation. Future work could explore integrating personalized background audio, such as music or nature-inspired soundscapes, to enhance presence and immersion.  Relaxation practices often emphasize the value of bodily grounding, which can be supported through modalities beyond sight and sound. For example, haptic textures corresponding to virtual environments, or even gentle airflow to simulate environmental cues, could enrich users’ sensory engagement. These modalities may help participants feel more anchored in the guided experience, reinforcing the imagery of safe and soothing spaces. Finally, this study mainly focused on respiration biofeedback. Expanding the range of biofeedback inputs such as heart rate variability, or electroencephalograph could offer a richer picture of stress and relaxation processes.

In addition, future work can improve the fidelity and realism of the generated environment maps, further supporting users’ ability to fully immerse themselves in their imagined safe spaces. In this study, we observed the mismatch between participants’ mental imagery and the system’s rendered 3D scenes or the lack of fine-grained detail may undermine user experience. With recent advances in 2D and 3D generative AI models, future systems could generate more photorealistic or spatially consistent environments, thereby deepening immersion and better supporting users’ ability to connect with the imagery in a meaningful way. The \name{} system in this study relied on a semi-manual workflow that requires facilitation by an art therapist. While this approach ensured guidance and safety, it limits scalability and long-term accessibility. Future work could explore more automatic or agentic workflows, where AI-driven agents streamline both the interpretation of user input and the generation of corresponding multi-modal outputs. Such automation would allow users to independently engage with the system outside of a supervised setting, while also improving the continuity of relaxation. 

\general{Regarding the generalizability and adaption, the biofeedback mechanism is largely universal and does not depend on cultural or therapeutic variation, whereas the Safe Place technique is highly dependent on users' narrative comfort and cultural norms surrounding emotional disclosure. The content people identify as “safe,” “distress,” or “wellness” is culturally shaped \cite{Gopalkrishnan_2018, OGUNDARE_2020}. Similarly, Yoshihara et al. \cite{Yoshihara_Acebes-de-Pablo_Honig_2024} highlighted the differences of how guided imagery was received and processed across culturally and linguistically diverse people. This implies that while our pipeline can technically generate any user-authored environment, the interpretive scaffolding including the imageries, metaphors, and sensory elaborations that feel natural would require cultural adaptation. For example, in collectivist communities where “safe spaces” may emphasize family, communal environments, or spiritual locations rather than solitary retreat \cite{Joshanloo_VanDeVliert_Jose_2021}. Based on the framework of \name, future works can ultilize prompts and image generation models that adapt to different cultures to enhance trust, engagement, and relaxation outcomes.}

\general{The personalization phase of \name {} relied on an art therapist to facilitate expressive detail and monitor psychological safety. This support influenced personalization depth: the therapist’s brief prompts helped participants articulate sensory elements and emotional associations that directly strengthened the resulting VR environments. Without such supervision, users may provide thinner, less emotionally resonant inputs, reducing the impact of personalization \cite{Reisner_Lith_Ashley_Mynard_2025}. On the other hand, scaling beyond therapist-supported sessions offers clear benefits—greater accessibility, lower cost, and potential deployment in everyday wellness or workplace settings. This motivates exploring AI-guided facilitation that can emulate key therapist functions: prompting users to elaborate imagery, ensuring grounding during reflection, and flagging moments of discomfort \cite{Olawade_Wada_Odetayo_David-Olawade_Asaolu_Eberhardt_2024}. Such AI agents would not replace clinical roles but could enable safe, independent use for non-clinical populations under established protocols, preserving personalization richness while reducing reliance on specialist supervision.}




\section{Conclusion}

We introduced ASafePlace, a VR biofeedback relaxation training system that incorporates user-led personalization enabled by \general{art-therapy-inspired activity}. Through \textit{The Safe Place} technique, the user's personal experiences as well as emotional and sensory memories are captured and transformed by AI into a user-authored 360° virtual environment and personalized relaxation guidance for biofeedback-assisted relaxation training. Our quantitative results revealed that personalization and biofeedback mutually reinforced each other, improving effectiveness of relaxation training. Besides, personalization via \general{art-therapy-inspired activity} strengthened user involvement and meaningfulness. Qualitative findings highlighted that \general{art-therapy-inspired activities} can be leveraged as a powerful tool for creating nuanced and in-depth personalization of VR relaxation experience. These findings point to the potential of autobiographical memory–driven and personal life experience–infused personalization in wellbeing and beyond. Finally, we discussed the insights and design implications including: combining verbal and non-verbal cues, embedding symbolic motifs, extending sensations through audio guidance, and organically linking biofeedback with personalized content. 


\begin{acks}
We sincerely appreciate the involvement of all the volunteers who participated in this study and the collaborating therapists for their generous support. We also greatly thank the reviewers for their invaluable feedback and suggestions. This work is supported by the Shenzhen STIC 2025 Grant (71): No. ZDCY20250901095359005.
\end{acks}

\bibliographystyle{ACM-Reference-Format}
\bibliography{sample-base}

\appendix

\section{Relaxation Training Guidance Scripts Template}

\textbf{This script was adapted and integrated from two meditation scripts in \cite{321}, and was used as the experimental material in this study.}

Welcome to this breath awareness meditation, a practice that invites you to connect with the simple yet profound rhythm of your breath. Find a quiet and comfortable place to sit or lie down. As you embark on this journey, allow your breath to be your guide, bringing you into the present moment with each inhale and exhale.

Close your eyes gently and take a few deep breaths. Inhale slowly through your nose, feeling your chest and abdomen expand, and then exhale through your mouth, releasing any tension.

Allow your breath to return to its natural rhythm. Observe the gentle flow of your breath, without trying to control it.

Please once again imagine and return to your Safe Place. This is a place where you feel safe and warm. The Safe Place belongs only to you, and no one else can enter without your permission.

Bring your attention to the sensation of breathing in. Notice the air as it enters your nostrils. Feel the breath filling you, and observe your chest rising and your belly expanding with each inhalation.

In this safe place, you are always calm. Notice the color of the sky at your favorite time of day. Let yourself experience this place at the perfect moment of the day, in the season and temperature you enjoy most, allowing your senses to become more vivid and alive.

Now shift your attention to the exhale. Feel the warmth of your breath as it leaves your body, and observe your chest and belly gently contracting. Follow the entire cycle of breath, from the inhale to the exhale.

Look around you, and allow yourself to see; and if not with your eyes, then with your heart. Each time you return to your safe place, you can make it even more beautiful.

Expand your awareness to the entire breathing cycle. Feel the seamless transition from inhalation to exhalation. Notice the pause between breaths, and in those quiet moments, cultivate a deeper sense of presence in the here and now.

Take a moment to release any tension or discomfort, and allow your body to settle into relaxation. As you continue to breathe naturally, bring awareness to different parts of your body. With each breath, imagine sending relaxation to any areas of tightness or unease, letting the breath cleanse and soothe them.

Let your breathing bring a sense of spaciousness and ease. Imagine that with every inhale, stress or tension gently melts away. Allow yourself to create a safe and peaceful place that belongs to you alone, always secure. Inhale safety, exhale fear. Inhale safety, exhale fear.

Take a few deep breaths to ground yourself. Give yourself this time to turn inward, letting go of any outside distractions. Imagine your breath as a gentle current of air. With each inhale, the current gathers around you, bringing fresh energy. With each exhale, the current flows outward, releasing all tension and pressure. Feel this rhythm, this calming energy, moving softly around you.

Now, shift your attention fully to your breathing. Notice the natural rhythm of inhaling and exhaling. Spend a few moments simply observing your breath, without trying to change it. Allow each breath to remind you to release any tension from your body.

Expand your awareness beyond your body. Feel your connection with the space around you, as though your breath extends outward and blends with the energy of the environment. With each exhale, let go of any remaining stress, and let soothing relaxation flow through every part of your body.

Here, you are completely safe. You can even sense it with taste, as if safety itself touches your lips. Allow yourself to be bathed in peace and protection. Take a moment to bring your attention to your whole body. Feel the ease and relaxation spreading through you, and enjoy this moment of being fully present, with a calm and peaceful body.

\section{Prompt for Personalized Relaxation Training Guidance Generation}

The Meditation Personalization Assistant specializes in extracting information from conversations between the visitor (client) and the therapist, and uses this information to personalize meditation guidance scripts. The goal is to enhance the effect of meditation, helping the visitor feel more relaxed and focused.

When the user does not understand how to use the system or what information to provide, explain the usage clearly and introduce the expected input and output. In all other cases, respond strictly according to the input–output rules defined below, without adding unrelated content.

\textbf{Input Rules}

The user should first provide the transcript of a conversation between the visitor and the therapist. The transcript has no annotations, so the model must distinguish between the therapist’s guidance/questions and the visitor’s self-expression. After receiving a segment of dialogue, the system should extract key information such as: what kind of scene the visitor drew that helps them feel relaxed; the name of this scene; what the visitor wishes to do in this scene; and details of the scene such as colors, textures, sounds, and smells. The model must not invent details not mentioned by the visitor. The output may begin with a summary of the key information extracted.

\textbf{Output Rules}

After extracting information, the system should follow the example sentences provided below. It must mimic their style and format to write personalized meditation guidance snippets. The tone should be gentle, friendly, and professional, ensuring that the visitor feels supported and cared for.

Each meditation snippet should mention only one sensory detail per sentence (for example, the first sentence about color, the second about texture, the third about sound, the fourth about smell, etc.). Beyond the four basic senses, the model should create as many additional snippets as possible, drawing on other details such as associations, memories, or activities the visitor wishes to do in this place. The four sample sentences alone are insufficient; more are required. The model may also incorporate the visitor’s associations or desires for this safe place.

When outputting personalized meditation snippets, the system must not include any text from the complete meditation template. Only the newly generated snippets should be provided at this stage. Each snippet should begin with a phrase that links it to the visitor’s earlier description, such as “As you mentioned before…” or “Just as you told me…”. Later, these snippets will be inserted into the full meditation template in place of placeholders.

\textbf{Example Meditation Snippets}

You are in the countryside, and it is winter. You walk on freshly fallen snow, hearing and feeling the crunch beneath your boots. The air is cold and crisp, and you can see your breath condense into mist. A church bell rings in the distance. Somewhere, a radio is playing “Jingle Bells.” Notice that sound.

You are on a beautiful, warm tropical beach. The sky is blue, the sun rests warmly on your skin. The sand beneath your feet is soft and warm. The vast ocean waves roll in endlessly, their rhythm never stopping.

You can hear the waves breaking, advancing and retreating. Imagine walking along the water’s edge, feeling the sand beneath your feet growing cooler. You walk a little faster as the sand grows hotter. When you reach the damp sand, feel the coolness beneath your feet. The water swirls around your ankles and, as it recedes, carries away some of the sand. The movement of sand and water leaves you feeling light and comfortable.

You are in a room from your childhood—one where you had joyful experiences. Notice what you see there, the sounds you hear, perhaps even a particular scent. Notice how it feels to be there.

The aroma of freshly ground coffee drifts toward you. A plate of your favorite food sits before you, carefully prepared with the best ingredients. You lean forward to smell it, enjoying the fragrance, then take a bite. Notice what it feels like as you eat, chew, and swallow.

You walk along a forest path. It is a beautiful day. On the way you meet a friendly person. You stop to have a brief conversation. Notice who this person is, what you talk about, and how you connect with each other.

\textbf{Integration into Template}

A complete meditation script template is provided. The model must ensure that the format and tone of the template remain unchanged. Only the content inside the curly braces { } is to be replaced. Each placeholder must be substituted with an appropriate personalized snippet. For fields labeled “choose another snippet,” the model should insert one of the remaining snippets that fits naturally; if none fit, the placeholder should be removed. Each substitution must match the sensory detail required by that placeholder (visual, tactile, auditory, olfactory, etc.). The original context, tone, and formatting of the template must be preserved.

At the end, the system outputs the fully personalized meditation script, including both the unchanged template text and the inserted personalized snippets.

\textbf{Complete Meditation Template}

Welcome to this breathing awareness meditation, a practice that allows you to connect with the simple yet profound rhythm of your breath. Find a quiet and comfortable place to sit or lie down. On this journey, let your breath be your guide, each inhalation and exhalation drawing you into the present moment.

Take a few deep breaths. Slowly inhale through your nose, feel your chest and belly expand, and then exhale through your mouth, releasing any tension.

Let your breathing return to its natural rhythm. Observe the gentle flow of breath without trying to control it.

Please once again imagine and return to {the name of the safe island}. This is a place where you feel safe and warm. {The name of the safe island} belongs only to you, and no one else can enter without your permission.

Bring your attention to the sensation of breathing in. Notice the air as it enters your nostrils. Feel the breath filling you, and observe your chest rising and your belly expanding with each inhalation.

Shift your attention to the exhale. Feel the warmth of your breath as it leaves your body, and observe your chest and belly gently contracting. Follow the entire cycle of breath, from the inhale to the exhale.

Look around you, and allow yourself to see; and if not with your eyes, then with your heart. Each time you return to your safe place, you can make it even more beautiful.

Expand your awareness to the entire breathing cycle. Feel the seamless transition from inhalation to exhalation. Notice the pause between breaths, and in those quiet moments, cultivate a deeper sense of presence in the here and now.

\{Meditation snippet related to visual details.\}

\{Meditation snippet related to tactile details.\} 

\{Meditation snippet related to auditory details.\}

As you continue to breathe naturally, bring awareness to different parts of your body. With each breath, imagine sending relaxation to areas of tension or discomfort, allowing the breath to cleanse and soothe them.

\{Choose another snippet.\}

Allow yourself to create a safe and peaceful place that belongs only to you, forever secure. Inhale safety, exhale fear. Inhale safety, exhale fear.

\{Choose another snippet.\}

Imagine your breath as a gentle current of air. With each inhale, the current gathers around you, bringing fresh energy. With each exhale, the current flows outward, releasing all tension and pressure. Feel the rhythm of this calming energy moving softly around you.

\{Meditation snippet related to olfactory details.\}

Expand your awareness beyond your body. Feel your connection with the space around you, as though your breath extends outward and blends with the energy of the environment.

\{Choose another snippet.\}

Here you are completely safe. You can even sense it with taste, as if safety itself touches your lips. Allow yourself to be bathed in peace and protection.

\{Choose another snippet.\}

Be grateful for this moment of mindful breathing, and slowly bring your awareness back to the present. Gently open your eyes, carrying with you the calm and centered energy of your breath.

May this breathing awareness meditation be a reminder of tranquility with each inhalation and exhalation. Whenever you seek peace during the day, return to this practice.

\end{document}